\documentclass[a4paper,fleqn, usenatbib]{mnras}

\usepackage[T1]{fontenc}
\usepackage{ae,aecompl}
\usepackage{longtable}

\usepackage{graphicx}	
\usepackage{amsmath}	
\usepackage{amssymb}	

\title[Reverberation Mapping of Mrk 590]{Changing look AGN Mrk 590: Broad line region and black hole mass from photometric reverberation mapping}

\author[Mandal et al.]{Amit Kumar Mandal$^{1,2}$\thanks{E-mail: amitkumar@iiap.res.in},
Malte Schramm$^{3}$,
Suvendu Rakshit$^{4}$,
C. S. Stalin$^{1}$,
\newauthor{Bovornpratch Vijarnwannaluk$^{5}$},
Wiphu Rujopakarn$^{6,7}$,
Saran Poshyachinda$^{7}$,
\newauthor{Vladimir V. Kouprianov$^{8,9}$},
Joshua B. Haislip$^{8}$, 
Daniel E. Reichart$^{8}$, 
\newauthor{Ram Sagar$^{1}$}
and Blesson Mathew$^{2}$
\\\\
$^{1}$Indian Institute of Astrophysics, Block II, Koramangala, Bangalore, 560 034, India\\
$^{2}$Department of Physics, CHRIST (Deemed to be University), Hosur Road, Bangalore 560 029, India\\
$^{3}$Graduate school of Science and Engineering, Saitama Univ. 255 Shimo-Okubo, Sakura-ku, Saitama City, Saitama 338-8570, JAPAN \\
$^{4}$Aryabhatta Research Institute of Observational Sciences, Manora Peak, Nainital 263002, India \\
$^{5}$ Astronomical Institute, Tohoku University, 6-3 Aramaki, Aoba-ku Sendai, Miyagi 980-8578, Japan\\
$^{6}$ Department of Physics, Faculty of Science, Chulalongkorn University, 254 Phayathai Road, Pathumwan, Bangkok 10330, Thailand \\
$^{7}$ National Astronomical Research Institute of Thailand, 260 Moo 4, T. Donkaew, A. Maerim, Chiangmai, 50180 Thailand \\
$^{8}$Department of Physics and Astronomy, University of North Carolina at Chapel Hill, Campus Box 3255, Chapel Hill, NC 27599-3255\\
$^{9}$ Central (Pulkovo) Observatory of the Russian Academy of Sciences
196140, 65/1 Pulkovskoye Ave., Saint Petersburg, Russia\\
}

\date{Accepted: 2021 October 5; Received: 2021 October 1; in original form: 2021 June 21}

\pubyear{2015}

\begin{document}
\label{firstpage}
\pagerange{\pageref{firstpage}--\pageref{lastpage}}
\maketitle

\begin{abstract}
We present the results of photometric reverberation mapping observations on the changing look active galactic nucleus Mrk 590 at z = 0.026. The observations were carried out from July to December, 2018 using broad band B, R and narrow band $\mathrm{H\alpha}$ and S II filters. B-band traces the continuum emission from the accretion disk, R-band encompasses both the continuum emission from the accretion disk and the redshifted $\mathrm{H\alpha}$ line from the broad line region (BLR), S II band contains the redshifted $\mathrm{H\alpha}$ emission and the $\mathrm{H\alpha}$ band traces the continuum emission underneath the S II band. All the light curves showed strong variation with a fractional root-mean-square variation of $0.132\pm0.001$ in $B$-band and $0.321\pm0.001$ in H$\alpha$ line. From cross-correlation function analysis, we obtained a delayed response of $\mathrm{H\alpha}$ line emission to the optical B-band continuum emission of $21.44^{+1.49}_{-2.11}$ days in the rest-frame of the source, corresponding to a linear size of the BLR of 0.018 pc. This is consistent with previous estimates using $\mathrm{H\beta}$. By combining the BLR size with the H$\alpha$ line full width at half maximum of $6478\pm240$ km s$^{-1}$ measured from a single-epoch spectrum obtained with the Subaru telescope, we derived a black hole mass of $1.96^{+0.15}_{-0.21}\times 10^8 M_{\odot}$.
\end{abstract}




\section{Introduction}

Active Galactic Nuclei (AGN) are powered by accretion of matter onto the supermassive black hole (SMBH; 10$^6$ $-$ 10$^{10}$  M$_{\odot}$) at the center of galaxies producing radiation covering the entire accessible electromagnetic spectrum. There is significant emission in the ultraviolet (UV)$-$optical blue continua from the accretion disk surrounding the SMBH in AGN \citep{1969Natur.223..690L,1964ApJ...140..796S}. The broad emission lines are produced due to the thermal re-ionization of UV/optical continuum in the broad line region (BLR) that lies outside the accretion disk. According to the unified scheme of AGN, an obscuring torus produces the infrared (IR) thermal emission from the hot dust characterized by the sublimation temperature $\mathrm{T_{sub}}$ at which the dust particle in radiative equilibrium with the disk UV radiation, sublimates. For graphite grains the sublimation temperature is $\mathrm{T_{sub} \sim 1700-2000}$ K \citep{2004ApJ...600L..35M, 2006ApJ...639...46S, 2007A&A...476..713K, 2014ApJ...788..159K}. The central part of AGN is too compact to resolve using conventional imaging techniques. Though \citet{2018Natur.563..657G} observed 3C 273 using GRAVITY and found a mean BLR size of 145 $\pm$ 35 days using the Pa${\alpha}$ line through interferometric technique, it is limited to only very bright and nearby AGN.

An indirect method to probe the BLR region in AGN is the technique of reverberation mapping \citep[RM;][]{1982ApJ...255..419B, 1993PASP..105..247P} that relies on the intrinsic flux variability properties of AGN. BLR RM is based on measuring the time delay between the UV/optical continuum from the accretion disk and the broad emission line produced in the BLR. As of today, BLR reverberation has provided the size of BLR in more than 100 AGN yielding a linear relation between the radius of the BLR ($\mathrm{R_{BLR}}$) and the continuum luminosity ($\mathrm{L_{5100{\AA}}}$) at 5100 \AA \citep{2015PASP..127...67B, 2017ApJ...851...21G, 2014ApJ...782...45D, 2014ApJ...793..108W} in the luminosity range between $10^{42} - 10^{46}$ erg s$^{-1}$. The $R_{BLR}-L$ relation and a single-epoch spectrum allow us to estimate black hole masses for a large number of AGN, but it is not clear if the same relation holds over a wide range of luminosity mainly due to the lack of measurements both at the low luminosity end ($L < 10^{42}$ erg s$^{-1}$) and the high luminosity end ($L > 10^{46}$ erg s$^{-1}$).



Recently, number of AGN have been discovered, which undergo a transition from Type 1 characterized by prominent broad-emission lines, to Type 1.8 or 1.9, with no or weak broad $\mathrm{H\alpha}$ and/or $\mathrm{H\beta}$, or vice versa \citep{2015toru.conf..A09L}. These objects are known as "changing-look" AGN, among which Mrk 590 is a well known object. Its optical spectrum obtained by \cite{1998ApJ...501...82P} showed strong broad H$\beta$ emission line but it disappeared between 2006 $-$ 2013 \citep{2014ApJ...796..134D} when the continuum luminosity decreased by a factor of 100 indicating that the black hole accretion rate has significantly decreased. The disappearance of broad emission lines in the UV/optical spectrum made Mrk 590 to appear as Seyfert 1.9$-$2,  where the only broad emission line visible in the optical spectrum is a weak component of $\mathrm{H\alpha}$ \citep{2014ApJ...796..134D}. However, recently, \citet{2019MNRAS.486..123R} noticed the reappearance of optical broad emission lines in Mrk 590 after $\sim 10$ yr of absence, while, the AGN optical continuum flux is still $\sim 10$ times lower than that observed during the most luminous state in the 1990s. During the period 1990 $-$ 1996 when Mrk 590 was in a bright state, $\mathrm{H\beta}$ BLR RM monitoring was carried out by \citet{1998ApJ...501...82P, 2004ApJ...613..682P} and \citet{2009ApJ...697..160B, 2013ApJ...767..149B}. The BLR lag based on $\mathrm{H\beta}$ line ranges from 19.5 days to 30 days \citep{2013ApJ...767..149B, 2011ApJ...735...80Z}.

Recently, dust reverberation mapping (DRM) study of Mrk 590 by \citet{2020MNRAS.491.4615K} suggests that the inner radius of the dust torus decreased after a drastic drop in the UV$-$optical luminosity. They found the innermost dust torus size of $\sim$ 32 light-days during the faint state in 2003 $-$ 2007, which is comparable to the $\mathrm{H\beta}$ BLR radius of $\sim$ 26 light-days obtained by RM observations during the bright state in 1990 $-$ 1996. We monitored Mrk 590 during the period July 2018 to December 2018 using broad $B$, $R$, narrow $\mathrm{H\alpha}$ and S II bands to estimate the size of the BLR based on the time delay between the $\mathrm{H\alpha}$ line and the continuum. Furthermore, to obtain spectral information and calculate black hole mass, we obtained a single-epoch spectrum using the Subaru telescope on October 27, 2018. Our black hole mass measurement using $\mathrm{H\alpha}$ is a new addition to the existing information on Mrk 590. The outline of the paper is as follows. In section \ref{sec:obs} we describe the observation and data reduction process. The time series analysis is presented  in section \ref{sec:analysis} and the results are discussed in section \ref{sec:discussion} with a summary in section \ref{sec:summary}.

\begin{figure*}
\resizebox{14cm}{8cm}{\includegraphics{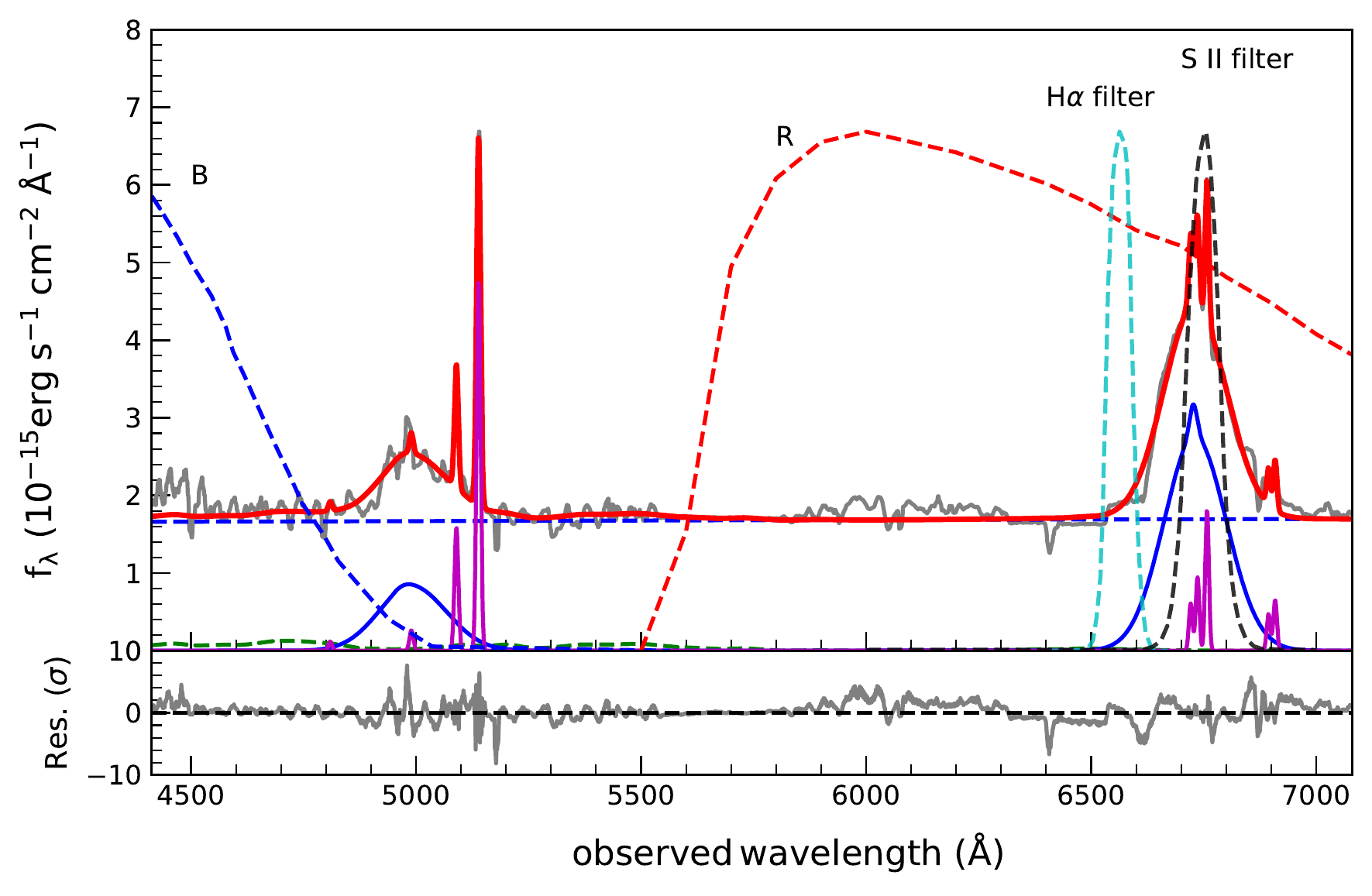}}
\caption{Subaru spectrum of Mrk 590. Top panel: The observed data (gray), best-fit model (red) and decomposed power-law continuum model (black dashed), broad line (blue), and narrow line (magenta) are shown. The broad $B$, $R$ and narrow H$\alpha$ and S II filters are over-plotted with limited x-axis for illustration. Bottom panel: Residual in the unit of flux uncertainty. The spectrum was smoothed by a 300 pixel box-car for visualization purpose.}
\label{fig:spec}
\end{figure*}

\section{Observation and data reduction}\label{sec:obs}

\subsection{Observations}

The photometric observations in the optical broad $B$ and $R$ filters as well as narrow $\mathrm{H\alpha}$ and S II filters were performed from July 2018 to December 2018 using the 60 cm robotic telescope PROMPT8 located at the Cerro Tololo Inter-American Observatory (CTIO) operated as part of Skynet \citep[see][for some introduction of the network]{2019ApJS..240...12M}. The observation period covers a total of 169 days with an attempted daily cadence. Our final data set consists of 115, 90, 86 and 83 epochs in $B,R,\mathrm{H\alpha}$ and SII, respectively, achieving an effective cadence of $\sim$1.5 days in the B-band and $\sim$2 days in SII. Gaps in the light-curves are mainly due to weather loss or moon distance which prevented observations. The B-band was given the highest priority for the Skynet scheduler as it defines our baseline for the AGN continuum which requires the best possible cadence. The CCD has an average readout noise and gain of 10.21 electrons and 2.2 electrons/ADU, respectively, with a pixel scale of 0.662 arcsec/pixel covering a FoV of $23'\times23'$. We used a 9-point dither pattern with a dither length of 50 arcsec to minimize the effect of bad pixel and cosmic rays. The typical total exposure times in $B$, $R$, $\mathrm{H\alpha}$ and S II bands are about 240s, 240s, 900s and 900s, respectively. Since calibration images were not available for each night, the calibration closest in time was assigned to the target observation epoch.   

\subsection{Optical spectrum}\label{sec:spectrum}

  \begin{table*}
  \caption{Spectral properties of Mrk 590. Columns are (1-2) FWHM and luminosity of H$\beta$ broad component, (3-4) FWHM and luminosity of [O III]$\lambda$5007, (5-6) FWHM and luminosity of H$\alpha$ broad component, (7) luminosity of H$\alpha$ narrow component, and (8) luminosity at 5100\AA. The units of FWHM and luminosity are km s$^{-1}$ and $\rm erg\;s^{-1}$, respectively. Line widths are not corrected for instrumental resolution.}
  \begin{center}
  \begin{tabular}{ccccccccc} \\ \hline \hline
  FWHM$(\mathrm{H\beta})$ & $\log L(\mathrm{H\beta})$ & FWHM([O III])  & $\log L$([O III]) &  FWHM$(\mathrm{H\alpha})$ & $\log L$ (H$\alpha$) & $\log L$(H$\alpha_n$) & $ \log \lambda L_{\lambda}$(5100) \\ 
  (1) & (2)       & (3)                  & (4)                   & (5)         & (6) & (7)  & (8)  \\ \hline
  $10390 \pm 1718$ & $41.38 \pm 0.01$ & $592 \pm 7$ & $40.88 \pm 0.01$ & $6478 \pm 240$ & $41.86\pm0.01$ & $40.13 \pm 0.08$ & $43.12 \pm 0.01$\\   
  \hline\hline
  \end{tabular}
  \label{Table:spec_fit}
  \end{center}
  \end{table*}
  
We obtained a single-epoch spectrum with an exposure time of 3600s using the  High Dispersion Spectrograph \citep[HDS;][]{10.1093/pasj/54.6.855} at the f/12.71 optical Nasmyth focus of the 8.2m Subaru telescope on October 27, 2018  under poor weather conditions. We used a long slit with 1 arcsec width and the data were sampled by two detectors (blue and red CCDs) with $4100 \times 2248$ pixels with a $2 \times 2$ binning resulting in a spectral resolution of R = 18,000 with a S/N of 10 in the continuum. The data was reduced using \textsc{iraf} \footnote{\textsc{iraf} is operated by the Association of Universities for Research in Astronomy, Inc., under cooperative agreement with the National Science Foundation} (Image Reduction and Analysis Facility) tools and a custom built application written in Python 2.7 making use of the available command language scripts in \textsc{iraf}  for overscan and non-linearty correction obtained from the HDS web site\footnote{https://www.subarutelescope.org/Observing/Instruments/HDS/index.html}.  The spectrum covers a wavelength range of 4354 $-$ 7138 {\AA} and shown in Fig. \ref{fig:spec}, where the $B$, $R$, H$\alpha$ and S II filter transmission curves are over-plotted.

To measure the emission line parameters, the Subaru spectrum was first corrected for Galactic extinction using the \citet{1998ApJ...500..525S} map and the Milky Way extinction law of \citet{1999PASP..111...63F} with $R_V=3.1$ and then moved to the rest-frame. A multi-component modeling was performed to estimate the emission line parameters. In the first step, we modeled the continuum, masking prominent emission lines, as a combination of a power-law ($f_{\lambda}\propto\lambda^{\alpha}$) and an Fe II template from \citet{1992ApJS...80..109B} that
accounts for the Fe II emission in AGN. The H$\alpha$ and H$\beta$ complexes of the best-fit continuum subtracted spectrum, was modeled using multiple Gaussians \citep[e.g.,][]{2020ApJS..249...17R}; upto three Gaussians for the broad lines and a single Gaussian for the narrow lines. The best-fit model is shown in Fig. \ref{fig:spec} and the parameters are given in Table \ref{Table:spec_fit}.

The broad B-band effectively traces the AGN continuum as the contribution of the higher order Balmer lines is negligible. The broad R-band contains the redshifted $\mathrm{H\alpha}$ line and the continuum flux from the accretion disk. The narrow S II band traces a large part of the redshifted H$\alpha$ (see Fig. \ref{fig:spec}) line and thus the S II narrow band is mostly dominated by the response of the emission line clouds in the BLR. The AGN continuum underneath the emission line is traced by the narrow H$\alpha$ band. Therefore, the pure emission line flux can be estimated by subtracting the narrow band H$\alpha$ flux from the narrow band S II flux. Also, H$\alpha$ emission line fluxes can be obtained from the R-band observations, however, continuum subtraction from observed R-band flux is a difficult task considering the larger accretion disk contribution compared to the narrow S II band. 

\subsection{Data reduction}\label{sec:reduc}

We used \textsc{iraf} to reduce all the broad $B$ and $R$, and narrow $\mathrm{H\alpha}$ and S II-band data. We followed the standard procedures for image reduction, such as bias and dark subtraction and flat-fielding.

\begin{figure}
\includegraphics[scale=0.5]{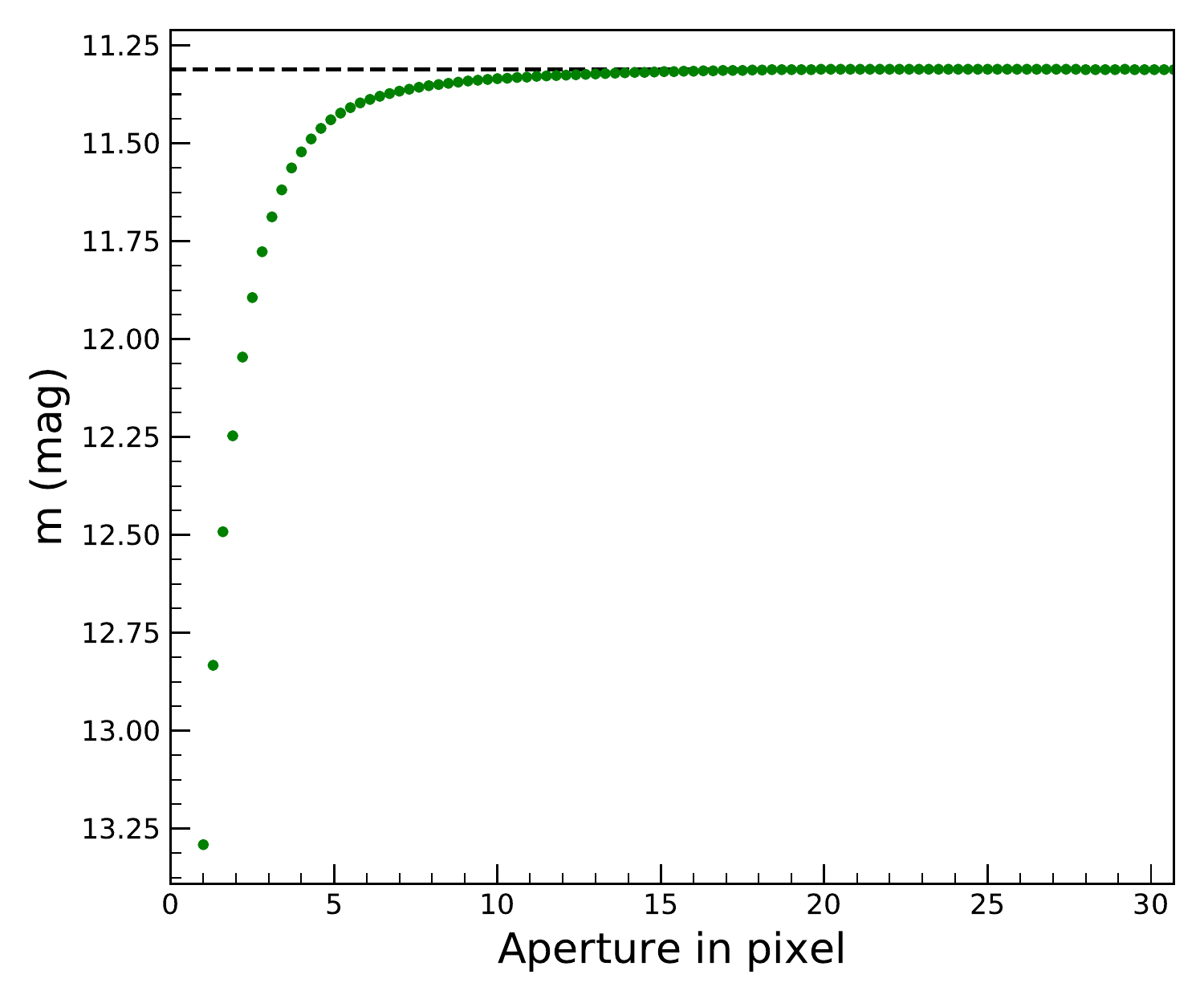}
\caption{Growth curve for a comparison star in S II-band. The dashed black line is the best fit line that uses the points with aperture sizes between 5 and 8 times the FWHM.}
\label{fig:fig-1}
\end{figure}

The {\it daofind} task available in \textsc{iraf} was used to detect the objects in the preprocessed images. Aperture photometry was performed using the {\it phot} task in \textsc{iraf}. The final instrumental magnitudes were obtained using the method of curve of growth \citep[COG;][]{2021MNRAS.501.3905M}. We obtained the instrumental magnitudes in several concentric apertures centered on few comparison stars with good signal to noise ratio (S/N) and, present on the same CCD frame.  The COG was produced by plotting the instrumental magnitudes obtained from different apertures against their corresponding aperture radii as shown in Fig. \ref{fig:fig-1}. The growth curve increases with aperture till the aperture is large enough to capture most of the flux from the source before gradually merging with the background. We fit a straight line based on the method of least squares to the portion of the growth curve with apertures between 5 and 8 times the FWHM of the comparison stars, where the COG merges with the background. The intercept of the fitted line (dashed black line in Fig. \ref{fig:fig-1}) was taken as the desired instrumental magnitude of the point source. This method was applied to each epoch to get the instrumental magnitudes of the comparison stars present in the CCD frames for all the bands.

\subsection{Correction for the host-galaxy contribution}

\begin{figure}
\includegraphics[scale=0.18]{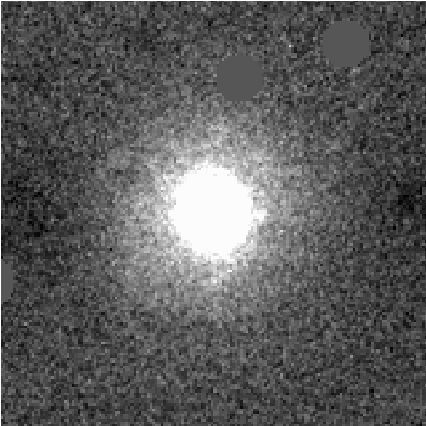}
\includegraphics[scale=0.18]{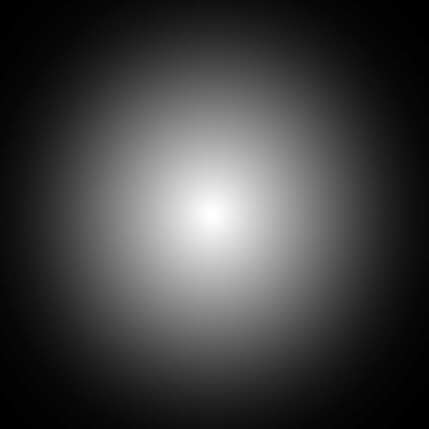}
\includegraphics[scale=0.18]{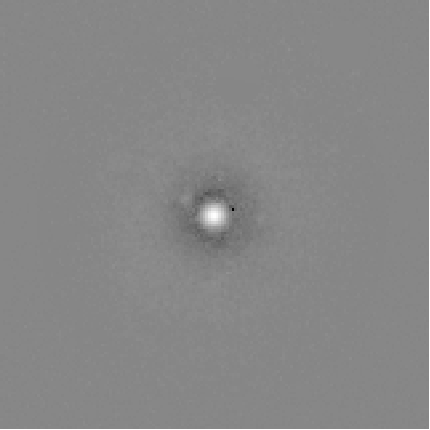}

\caption{Observed S II-band image of Mrk 590 (left). Modeled galaxy image (middle) and residual image containing the AGN at the center (right) obtained from GALFIT. The modeled and residual images are displayed in logarithmic scale. The field of view of each image is $1.8' \times 1.8'$.}
\label{fig:fig-2}
\end{figure}

The host galaxy is prominently seen in the observed images of Mrk 590 in all the filters (see the left panel of Fig. \ref{fig:fig-2} for S II filter). To remove the host-galaxy contribution from the total flux, we used the two dimensional image-decomposition program \textsc{galfit} developed by \citet{2002AJ....124..266P,2010AJ....139.2097P}. We combined several frames of different epochs with good seeing condition covering the whole span of our monitoring period to obtain a single deep image with high S/N for each of the four bands separately. Then we applied the \textsc{galfit} to decompose the combined images and extract a host galaxy model.  

\citet{2017ApJS..232...21K} performed \textsc{galfit} on Mrk 590 to model the galaxy and found that the host galaxy has a pseudo bulge with S\'ersic index n = 1.04 and a disk. An exponential disk profile is simply a S\'ersic profile with an index of n = 1. Following \citet{2017ApJS..232...21K}, we used exponential disk profile to model the underlying galaxy of Mrk 590. We generated PSF image from point sources in the combined images. The modeled galaxy image and the residual image that contains the AGN for the S II-band are shown in Fig. \ref{fig:fig-2} (middle) and (right), respectively.

\begin{figure}
\resizebox{9cm}{7cm}{\includegraphics{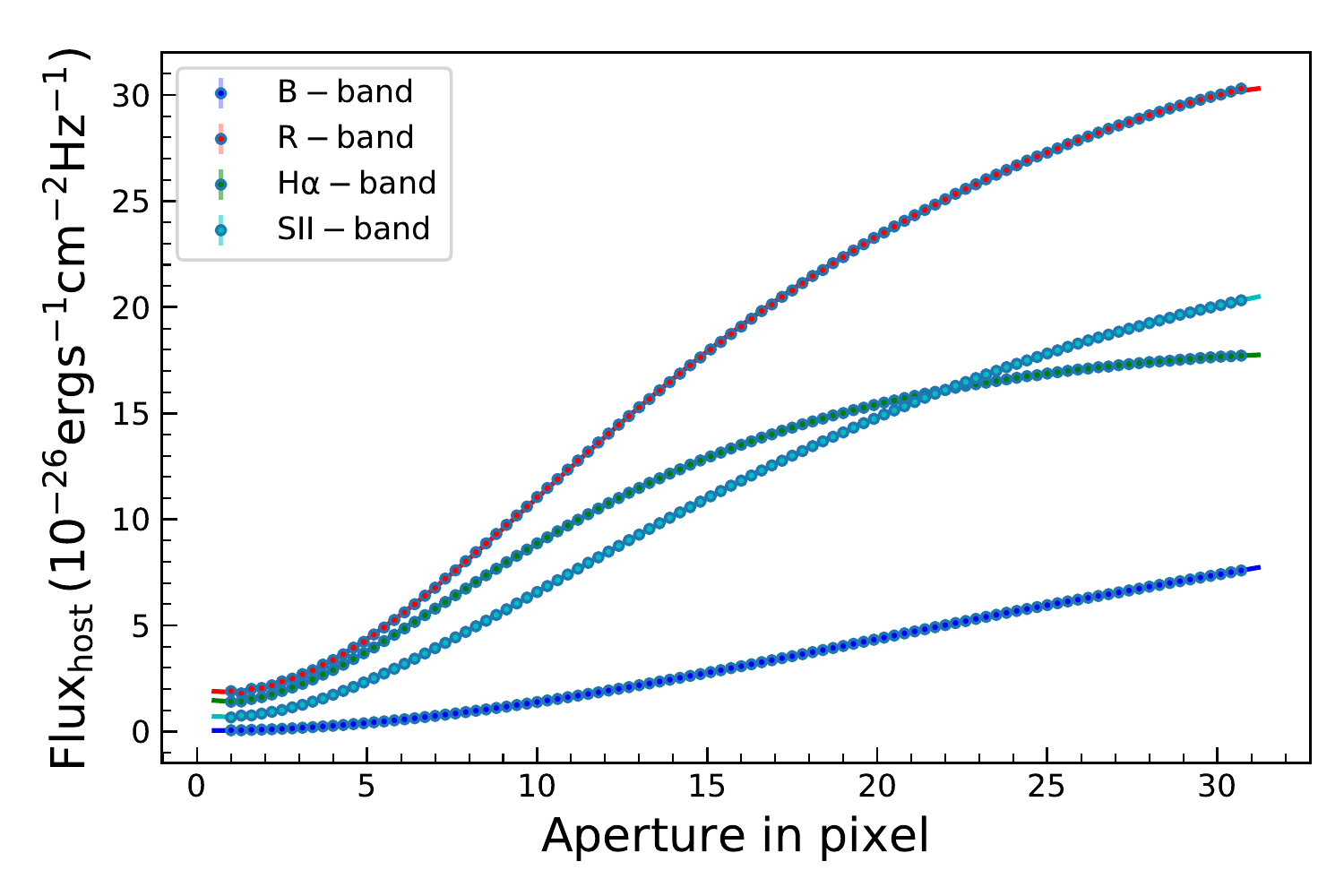}}
\caption{Host-galaxy flux as a function of aperture sizes in $B$, $R$, $\mathrm{H\alpha}$ and S II bands, respectively. The points with errorbar are the galaxy-flux values and the solid lines are the best polynomial fits. The original $B$-band fluxes are multiplied by 10 for the purpose of presentation.}
\label{fig:fig-galx}
\end{figure}

The AGN flux, ideally, can be obtained from the photometry of the residual image that contains only the AGN at the center. But practically, it is very difficult due to poor S/N in many epochs of data. Therefore, we carried out aperture photometry on the modeled galaxy image at different concentric aperture sizes and the estimated galaxy-fluxes are plotted against the aperture in Fig. \ref{fig:fig-galx}. Here, the solid lines show the best-fit curves representing the functional forms of the galaxy light distribution. We performed aperture photometry of the target Mrk 590 at an aperture equal to the aperture radius used for the comparison stars at each individual epoch. We obtained the host-galaxy fluxes from the best fit curve at the aperture used for the photometry of Mrk 590. Then we subtracted the host-galaxy fluxes from the observed total fluxes in those filters to get the host free AGN flux.

\subsection{Flux Calibration}

The instrumental magnitudes obtained from aperture photometry in $B$, $R$, $\mathrm{H\alpha}$ and S II bands were converted to fluxes through differential photometry using the following steps:

1. We collected $u$ and $g$-band apparent magnitudes from the SDSS database for a few comparison stars with the best S/N present in the same observed frame. The $B$-band apparent magnitudes of those comparison stars were obtained from SDSS $u$ and $g$- magnitudes using the relation\footnote{https://www.sdss.org/dr12/algorithms/sdssUBVRITransform/ \\ $\#$Lupton2005} given below
\begin{equation}
B = u - 0.8116(u-g) + 0.1313
\end{equation}
Then we converted the $B$-band instrumental magnitude to the standard $B$-band magnitude of Mrk 590 using differential photometry of the comparison stars. We collected the $R$-band standard magnitudes of the comparison stars from \citet{2017yCat.1340....0Z} and the $R$-band magnitude of Mrk 590 was obtained using differential photometry of the comparison stars.

2. Narrow $\mathrm{H\alpha}$ and S II -filters have a mean wavelength of $\sim$ 6563 {\AA} and $\sim$ 6719 {\AA} \citep{2011BaltA..20..459A}, respectively. The SDSS $r$-band has wavelength coverage from 5380 {\AA} to 7230 {\AA} with the central wavelength at 6166 {\AA}. Therefore, both the narrow $\mathrm{H\alpha}$ and S II band completely fall on the SDSS $r$-band. Hence, we collected the $r$-band apparent magnitudes of the comparison stars from the SDSS database and obtained the standard magnitude of  Mrk 590 in both $\mathrm{H\alpha}$ and S II filters using differential photometry of the comparison stars. 

The apparent magnitudes were corrected for Galactic extinction taken from the NASA/IPAC Extragalactic data base (NED)\footnote{http://nedwww.ipac.caltech.edu} for the broad $B$ and $R$ bands and for the narrow $\mathrm{H\alpha}$ and S II bands, Galactic extinction in r-band was used. The extinction corrected $B$ and $R$-band magnitudes of Mrk 590 were converted to fluxes using the conversion factors given in \citet{1998A&A...333..231B}. Similarly, to convert the $\mathrm{H\alpha}$ and S II magnitudes of Mrk 590, we used the SDSS r-band conversion factor\footnote{https://www.sdss.org/dr12/algorithms/fluxcal/}.

\begin{figure}
\resizebox{8cm}{8cm}{\includegraphics{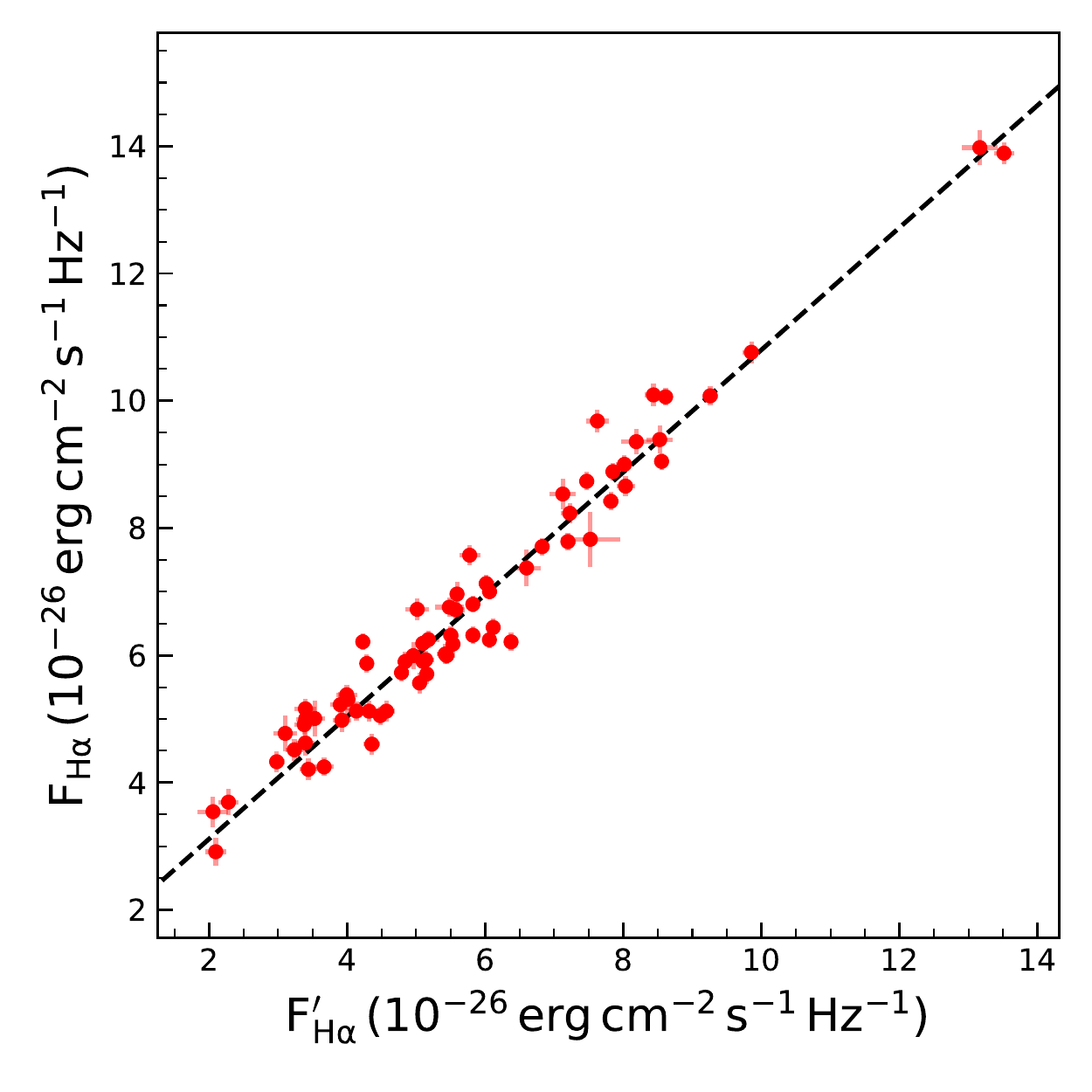}}
\resizebox{8cm}{8cm}{\includegraphics{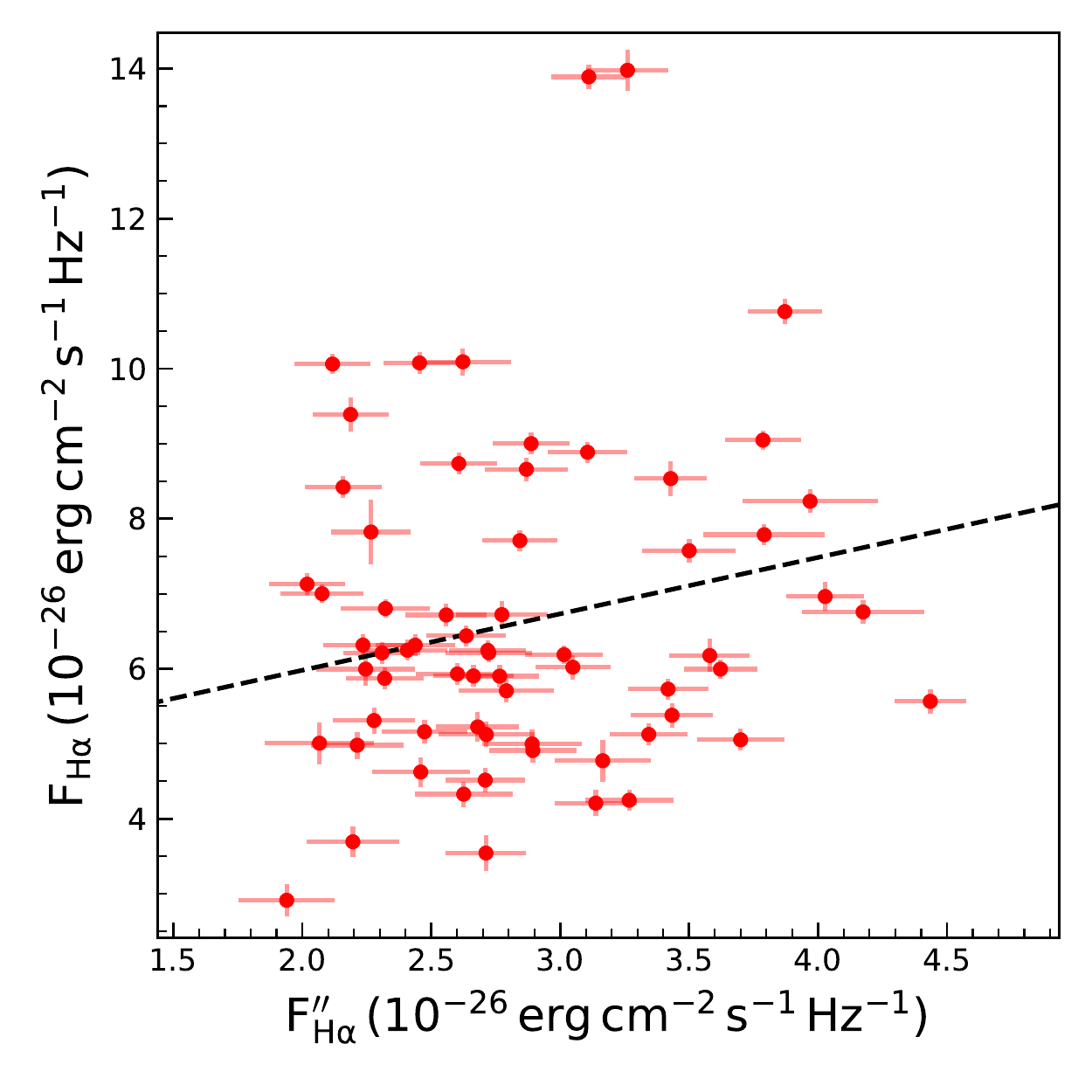}}
\caption{The $\mathrm{H\alpha}$ fluxes ($\mathrm{F_{H\alpha}}$) evaluated using the continuum observations
done in the narrow $\mathrm{H\alpha}$ filter as a function of $\mathrm{H\alpha}$ fluxes ($\mathrm{F_{H\alpha}^{\prime}}$) evaluated using broad B-band (top) and that ($\mathrm{F_{H\alpha}^{\prime\prime}}$) obtained from R-band using PL spectrum of accretion disk (bottom). The dashed lines represent the best linear fits. Near simultaneous observations between the narrow $\mathrm{H\alpha}$ filter and the broad $B$ and $R$-bands were used in both the plots.}
\label{fig:fig-Fhlpcor}
\end{figure}

\subsection{Generation of $\mathrm{H\alpha}$ emission line light curve}\label{sec:genlc}

The accumulated data on Mrk 590, enables generation of $\mathrm{H\alpha}$ emission line light curve by three methods.

\subsubsection{Method 1}

The observed flux in the narrow S II-band ($F_{\mathrm{S \, II,tot}}$) contains both $\mathrm{H\alpha}$ line ($F_{\mathrm{H\alpha}}$) and continuum fluxes ($F_{\mathrm{cont}}$). The variable continuum originates from the accretion disk whose contribution to the observed S II-band fluxes would make the derived lag between the optical/UV continuum and $\mathrm{H\alpha}$ line shorter than the actual lag \citep{2014ApJ...788..159K}. The accretion disk contributions to the observed line fluxes need to be removed first to get the actual time lag between the  $B$ and $\mathrm{H\alpha}$ light curves. The observations in the narrow $\mathrm{H\alpha}$-band contains the continuum flux ($F_{\mathrm{cont, H\alpha}}$) from the accretion disk. Hence, $F_{\mathrm{H\alpha}}$ can be obtained from the observed $F_{\mathrm{S \, II,tot}}$ as follows

\begin{equation}
F_{\mathrm{H\alpha}}(t) = F_{\mathrm{S \, II,tot}}(t) - F_{\mathrm{cont, H\alpha}}(t)
\label{eq:hlp_ori}
\end{equation}

\subsubsection{Method 2}

The  continuum contribution from the accretion disk to the observed S II-band flux can also be estimated from the observed B-band data. This is done by considering a power-law (PL) spectrum of the accretion disk given
by \citet{2014ApJ...788..159K}

\begin{equation}
F_{\mathrm{S \, II,cont}}(t) = F_B(t) \big(\frac{\nu_{\mathrm{S \, II}}}{\nu_{B}}\big)^{\alpha}
\label{eq:cont_sub}
\end{equation}   

where $F_{\mathrm{S \, II,cont}}$(t) and $F_B$(t) represent the continuum flux of the S II-band from the accretion disk and the $B$-band flux at time t, respectively. The $\nu_{B}$ and $\nu_{\mathrm{S \, II}}$ are the effective frequencies of the $B$ and S II bands while $\alpha$ is the power-law index. According to \citet{2006ApJ...652L..13T}, the power-law index of the accretion-disk component of the flux in the optical and near-infrared may vary from $-$0.1 to +0.4. \citet{2008Natur.454..492K} found $\alpha = 0.44 \pm 0.11$ from spectropolarimetric observations. Using the quasar composite spectrum observed by the Sloan Digital Sky Survey, \citet{2005ApJ...633..638W} obtained a power-law index of $\alpha$ = 0. We calculated the continuum contribution in the S II-band flux for $\alpha$ = 1/3 as prescribed by standard accretion disk model \citep{1973A&A....24..337S}. Finally, we subtracted the $F_{\mathrm{S \, II,cont}}$ from the observed S II-flux to get the $\mathrm{H\alpha}$ line flux ($F_{\mathrm{H\alpha}}^{\prime}$) as

\begin{equation}
F_{\mathrm{H\alpha}}^{\prime}(t) = F_{\mathrm{S \, II,tot}}(t) - F_{\mathrm{S \, II,cont}}(t)
\label{eq:cont_pow}
\end{equation}

The $\mathrm{H\alpha}$ fluxes evaluated using the continuum observations done in the narrow $\mathrm{H\alpha}$ filter (Equation \ref{eq:hlp_ori}) and broad B-band (Equation \ref{eq:cont_pow}) are found to follow a linear trend, however $F_{\mathrm{H\alpha}}$ is always larger than $F_{\mathrm{H\alpha}}^{\prime}$. This is shown in Fig. \ref{fig:fig-Fhlpcor} (top) that includes those points which have near simultaneous observations in B and  $\mathrm{H\alpha}$ filters. From linear least squares fit to the data, we found
\begin{equation}
F_{\mathrm{H\alpha}} = (0.96 \pm 0.03) F_{\mathrm{H\alpha}}^{\prime} + 1.19 \pm 0.17
\end{equation}

\subsubsection{Method 3}

The observed $R$-band fluxes contain continuum contribution from the accretion disk and the redshifted $\mathrm{H\alpha}$ line flux from the BLR. To get a lag between $B$ and the $\mathrm{H\alpha}$ line flux derived using the observations in the broad $R$-band, it is necessary to subtract the continuum contribution from the observed $R$-band flux. However, such decomposition is difficult to perform as the AGN continuum is the major contributor to the total flux. Assuming a PL spectrum of accretion disk, we measured the AGN continuum contribution to the observed host-corrected $R$-band fluxes from the near simultaneous $B$-band fluxes as

\begin{equation}
F_{\mathrm{R,cont}}(t) = F_B(t)\big(\frac{\nu_{R}}{\nu_{B}}\big)^{\alpha}
\label{eq:cont_R}
\end{equation}

We used $\alpha = 1/3$ considering a standard accretion disk. From the observed $R$-band fluxes the
derived $R$-band continuum was subtracted to get the $\mathrm{H\alpha}$ line fluxes as

\begin{equation}
F_{\mathrm{H\alpha}}^{\prime\prime}(t) = F_R(t) - F_{\mathrm{R,cont}}(t)
\label{eq:hlp_R}
\end{equation}

The $\mathrm{H\alpha}$ fluxes ($\mathrm{F_{H\alpha}}$) evaluated using the continuum observations done in the narrow $\mathrm{H\alpha}$ filter (Equation \ref{eq:hlp_ori}) are plotted as a function of the $\mathrm{H\alpha}$ line fluxes ($\mathrm{F_{H\alpha}^{\prime\prime}}$) derived from Equation \ref{eq:hlp_R} in Fig \ref{fig:fig-Fhlpcor} (bottom). We could not find any strong correlation between $\mathrm{F_{H\alpha}^{\prime\prime}}$ and $\mathrm{F_{H\alpha}}$. From linear least squares fit to the data points, we found a Pearson correlation coefficient of r $=$ 0.152 with a p value of 0.228. This indicates that it is not possible to derive $\mathrm{H\alpha}$ line fluxes from broad R-band observations. The final light curve data except that of $\mathrm{F_{H\alpha}^{\prime\prime}}$ are given in Table \ref{tab:table-lcdata}.

\begin{table*}
\caption{Results of photometry. The fluxes and their uncertainties in 
different photometric bands are in units of $\mathrm{10^{-26} \, erg \, s^{-1} \, cm^{-2} \, Hz^{-1}}$. MJD in days is given as MJD $-$ 58310.0.}

\resizebox{19cm}{!}{
\begin{tabular}{cccccccccccccccccc} \hline

\multicolumn{3}{c}{$B$-band} &
\multicolumn{3}{c}{$R$-band} &
\multicolumn{3}{c|}{$S \, II$-band} &
\multicolumn{3}{c}{$\mathrm{H\alpha}$-band} &
\multicolumn{3}{c}{$\mathrm{H\alpha}$ line} &
\multicolumn{3}{c}{$\mathrm{{H\alpha}^\prime}$ line} \\

MJD & F  & $\sigma$ & MJD & F  & $\sigma$ &  MJD & F  & $\sigma$ & MJD & F  & $\sigma$ & MJD & F  & $\sigma$ & MJD & F  & $\sigma$  \\

\hline
4	&	2.891	&	0.075	&	3	&	4.282	&	0.146	&	10	&	6.624	&	0.121	&	3	&	1.191	&	0.111	&	10	&	5.122	&	0.170	&	10	&	4.318	&	0.164	\\
5	&	2.554	&	0.045	&	10	&	5.056	&	0.150	&	13	&	7.874	&	0.110	&	10	&	1.502	&	0.120	&	13	&	6.756	&	0.155	&	13	&	5.481	&	0.207	\\
10	&	2.663	&	0.127	&	13	&	6.606	&	0.158	&	14	&	6.083	&	0.104	&	12	&	1.237	&	0.394	&	14	&	4.604	&	0.165	&	14	&	4.359	&	0.120	\\
12	&	2.764	&	0.202	&	14	&	4.531	&	0.144	&	18	&	6.281	&	0.109	&	13	&	1.118	&	0.110	&	18	&	5.123	&	0.163	&	18	&	4.571	&	0.127	\\
26	&	1.919	&	0.109	&	18	&	5.024	&	0.143	&	25	&	6.172	&	0.099	&	14	&	1.479	&	0.128	&	25	&	5.052	&	0.145	&	25	&	4.480	&	0.113	\\
29	&	1.990	&	0.069	&	25	&	5.388	&	0.139	&	26	&	5.649	&	0.106	&	15	&	0.689	&	0.566	&	26	&	4.247	&	0.144	&	26	&	3.668	&	0.140	\\
30	&	1.975	&	0.074	&	26	&	4.956	&	0.142	&	28	&	4.198	&	0.205	&	18	&	1.158	&	0.121	&	28	&	3.543	&	0.239	&	28	&	2.060	&	0.220	\\
31	&	1.955	&	0.062	&	28	&	4.463	&	0.143	&	30	&	15.234	&	0.250	&	19	&	0.615	&	0.159	&	30	&	13.978	&	0.272	&	30	&	13.162	&	0.262	\\
32	&	2.287	&	0.106	&	29	&	4.024	&	0.144	&	31	&	5.224	&	0.109	&	22	&	1.227	&	0.147	&	31	&	4.515	&	0.168	&	31	&	3.238	&	0.131	\\
33	&	2.469	&	0.093	&	30	&	4.999	&	0.146	&	32	&	7.204	&	0.103	&	25	&	1.120	&	0.106	&	32	&	5.931	&	0.142	&	32	&	5.139	&	0.119	\\
34	&	2.393	&	0.093	&	31	&	4.429	&	0.143	&	34	&	6.079	&	0.095	&	26	&	1.402	&	0.098	&	34	&	5.380	&	0.161	&	34	&	3.996	&	0.148	\\
35	&	2.293	&	0.084	&	32	&	4.614	&	0.132	&	37	&	6.370	&	0.102	&	28	&	0.655	&	0.122	&	37	&	5.125	&	0.149	&	37	&	4.136	&	0.166	\\
36	&	2.629	&	0.069	&	34	&	5.538	&	0.136	&	39	&	9.815	&	0.133	&	29	&	0.809	&	0.117	&	39	&	9.682	&	0.174	&	39	&	7.625	&	0.168	\\
37	&	2.384	&	0.068	&	37	&	5.441	&	0.139	&	40	&	7.823	&	0.113	&	30	&	1.256	&	0.109	&	40	&	7.573	&	0.153	&	40	&	5.775	&	0.151	\\
40	&	2.406	&	0.131	&	40	&	5.617	&	0.140	&	42	&	6.927	&	0.127	&	31	&	0.708	&	0.128	&	42	&	6.722	&	0.176	&	42	&	5.017	&	0.174	\\
41	&	2.580	&	0.151	&	42	&	4.999	&	0.146	&	43	&	4.886	&	0.161	&	32	&	1.273	&	0.098	&	43	&	4.773	&	0.280	&	43	&	3.105	&	0.177	\\
42	&	2.529	&	0.119	&	43	&	5.390	&	0.155	&	45	&	5.773	&	0.121	&	34	&	0.699	&	0.129	&	45	&	4.980	&	0.180	&	45	&	3.927	&	0.129	\\
47	&	2.365	&	0.116	&	45	&	4.293	&	0.149	&	46	&	3.937	&	0.142	&	35	&	0.475	&	0.344	&	46	&	3.693	&	0.208	&	46	&	2.283	&	0.151	\\
49	&	2.206	&	0.137	&	46	&	4.276	&	0.149	&	47	&	3.655	&	0.142	&	36	&	0.684	&	0.196	&	47	&	2.914	&	0.220	&	47	&	2.099	&	0.151	\\
55	&	2.056	&	0.085	&	47	&	4.020	&	0.155	&	48	&	5.191	&	0.117	&	37	&	1.245	&	0.109	&	48	&	4.623	&	0.197	&	48	&	3.397	&	0.129	\\
56	&	2.132	&	0.052	&	48	&	4.400	&	0.146	&	49	&	5.567	&	0.112	&	39	&	0.133	&	0.112	&	49	&	4.998	&	0.194	&	49	&	3.401	&	0.135	\\
57	&	1.910	&	0.061	&	49	&	4.832	&	0.149	&	50	&	5.094	&	0.101	&	40	&	0.250	&	0.102	&	50	&	4.328	&	0.170	&	50	&	2.982	&	0.117	\\
58	&	1.797	&	0.057	&	50	&	4.567	&	0.147	&	52	&	5.677	&	0.085	&	41	&	0.485	&	0.130	&	52	&	4.911	&	0.160	&	52	&	3.381	&	0.140	\\
59	&	2.072	&	0.064	&	52	&	4.703	&	0.151	&	53	&	5.753	&	0.103	&	42	&	0.205	&	0.121	&	53	&	5.160	&	0.154	&	53	&	3.397	&	0.155	\\
60	&	2.501	&	0.086	&	53	&	4.283	&	0.149	&	54	&	7.768	&	0.120	&	44	&	0.112	&	0.229	&	54	&	6.717	&	0.155	&	54	&	5.572	&	0.137	\\
62	&	2.439	&	0.067	&	54	&	4.367	&	0.139	&	56	&	7.199	&	0.096	&	45	&	0.792	&	0.133	&	56	&	6.247	&	0.138	&	56	&	5.178	&	0.154	\\
63	&	2.652	&	0.129	&	55	&	4.548	&	0.138	&	57	&	7.017	&	0.094	&	46	&	0.244	&	0.152	&	57	&	5.901	&	0.153	&	57	&	4.841	&	0.110	\\
65	&	2.721	&	0.135	&	56	&	4.283	&	0.149	&	58	&	7.354	&	0.085	&	47	&	0.741	&	0.168	&	58	&	6.020	&	0.162	&	58	&	5.422	&	0.121	\\
68	&	2.536	&	0.077	&	57	&	4.446	&	0.143	&	60	&	8.145	&	0.082	&	48	&	0.569	&	0.159	&	60	&	7.001	&	0.121	&	60	&	6.063	&	0.104	\\
69	&	2.334	&	0.139	&	58	&	4.630	&	0.137	&	61	&	9.933	&	0.119	&	52	&	0.766	&	0.136	&	61	&	8.659	&	0.157	&	61	&	8.032	&	0.133	\\
72	&	2.513	&	0.068	&	60	&	4.276	&	0.142	&	62	&	9.577	&	0.416	&	53	&	0.594	&	0.115	&	62	&	7.823	&	0.432	&	62	&	7.523	&	0.431	\\
73	&	2.231	&	0.099	&	61	&	5.069	&	0.141	&	63	&	9.089	&	0.101	&	54	&	1.051	&	0.099	&	63	&	7.787	&	0.139	&	63	&	7.198	&	0.108	\\
74	&	2.404	&	0.074	&	62	&	4.412	&	0.143	&	65	&	10.365	&	0.118	&	55	&	0.909	&	0.101	&	65	&	10.092	&	0.181	&	65	&	8.437	&	0.123	\\
75	&	2.195	&	0.068	&	63	&	6.124	&	0.206	&	66	&	10.264	&	0.168	&	56	&	0.952	&	0.099	&	66	&	9.357	&	0.199	&	66	&	8.188	&	0.221	\\
81	&	2.373	&	0.130	&	65	&	5.016	&	0.146	&	68	&	7.439	&	0.106	&	57	&	1.116	&	0.121	&	68	&	6.314	&	0.140	&	68	&	5.503	&	0.114	\\
83	&	2.184	&	0.045	&	68	&	4.668	&	0.139	&	69	&	7.062	&	0.104	&	58	&	1.334	&	0.138	&	69	&	5.706	&	0.157	&	69	&	5.154	&	0.110	\\
84	&	2.226	&	0.042	&	69	&	4.844	&	0.138	&	70	&	6.726	&	0.113	&	59	&	1.435	&	0.102	&	70	&	5.995	&	0.218	&	70	&	4.960	&	0.141	\\
85	&	2.397	&	0.165	&	70	&	4.299	&	0.146	&	73	&	6.690	&	0.076	&	60	&	1.144	&	0.090	&	73	&	6.213	&	0.129	&	73	&	4.229	&	0.088	\\
86	&	2.235	&	0.048	&	73	&	4.686	&	0.144	&	74	&	6.672	&	0.084	&	61	&	1.274	&	0.103	&	74	&	5.872	&	0.142	&	74	&	4.286	&	0.093	\\
87	&	2.203	&	0.041	&	74	&	4.434	&	0.136	&	75	&	6.238	&	0.088	&	62	&	1.755	&	0.115	&	75	&	5.310	&	0.177	&	75	&	4.013	&	0.099	\\
91	&	2.039	&	0.099	&	75	&	4.209	&	0.149	&	80	&	5.541	&	0.150	&	63	&	1.303	&	0.095	&	80	&	5.007	&	0.278	&	80	&	3.532	&	0.153	\\
93	&	2.842	&	0.051	&	80	&	4.154	&	0.178	&	83	&	8.067	&	0.088	&	65	&	0.273	&	0.138	&	83	&	6.802	&	0.126	&	83	&	5.822	&	0.095	\\
94	&	2.756	&	0.044	&	83	&	4.243	&	0.169	&	84	&	9.904	&	0.102	&	66	&	0.907	&	0.106	&	84	&	8.735	&	0.145	&	84	&	7.471	&	0.108	\\
95	&	2.569	&	0.053	&	84	&	4.565	&	0.144	&	85	&	8.939	&	0.199	&	68	&	1.124	&	0.092	&	85	&	7.373	&	0.290	&	85	&	6.600	&	0.203	\\
96	&	2.320	&	0.035	&	86	&	4.981	&	0.146	&	86	&	7.210	&	0.075	&	69	&	1.356	&	0.118	&	86	&	6.188	&	0.119	&	86	&	5.097	&	0.079	\\
97	&	2.593	&	0.041	&	87	&	3.956	&	0.144	&	87	&	8.088	&	0.109	&	70	&	0.730	&	0.186	&	87	&	7.128	&	0.143	&	87	&	6.016	&	0.114	\\
98	&	2.810	&	0.042	&	88	&	5.367	&	0.136	&	88	&	9.279	&	0.185	&	72	&	0.169	&	0.136	&	88	&	8.535	&	0.235	&	88	&	7.124	&	0.188	\\
99	&	2.702	&	0.045	&	91	&	5.763	&	0.247	&	91	&	9.353	&	0.119	&	73	&	0.477	&	0.104	&	91	&	8.232	&	0.157	&	91	&	7.227	&	0.125	\\
100	&	2.441	&	0.029	&	93	&	4.617	&	0.140	&	93	&	10.669	&	0.100	&	74	&	0.800	&	0.115	&	93	&	10.062	&	0.137	&	93	&	8.613	&	0.108	\\
101	&	2.393	&	0.038	&	94	&	4.582	&	0.144	&	94	&	9.811	&	0.096	&	75	&	0.929	&	0.153	&	94	&	8.421	&	0.144	&	94	&	7.821	&	0.115	\\
102	&	2.489	&	0.039	&	96	&	4.929	&	0.146	&	96	&	10.013	&	0.112	&	80	&	0.534	&	0.234	&	96	&	9.001	&	0.145	&	96	&	8.017	&	0.120	\\
103	&	2.456	&	0.041	&	97	&	4.735	&	0.136	&	97	&	10.859	&	0.107	&	83	&	1.265	&	0.090	&	97	&	10.077	&	0.148	&	97	&	9.258	&	0.110	\\
104	&	2.374	&	0.046	&	98	&	5.315	&	0.141	&	98	&	8.834	&	0.095	&	84	&	1.169	&	0.104	&	98	&	7.708	&	0.139	&	98	&	6.825	&	0.101	\\
106	&	2.298	&	0.074	&	99	&	4.686	&	0.144	&	99	&	8.077	&	0.104	&	85	&	1.567	&	0.211	&	99	&	6.211	&	0.142	&	99	&	6.375	&	0.106	\\
107	&	2.304	&	0.052	&	100	&	4.783	&	0.152	&	100	&	7.780	&	0.108	&	86	&	1.022	&	0.093	&	100	&	6.439	&	0.142	&	100	&	6.116	&	0.110	\\
109	&	1.849	&	0.027	&	101	&	4.824	&	0.145	&	101	&	7.877	&	0.087	&	87	&	0.960	&	0.093	&	101	&	6.244	&	0.131	&	101	&	6.061	&	0.090	\\
112	&	2.320	&	0.041	&	102	&	4.424	&	0.150	&	102	&	7.579	&	0.085	&	88	&	0.744	&	0.145	&	102	&	6.316	&	0.142	&	102	&	5.823	&	0.089	\\
113	&	1.965	&	0.027	&	103	&	4.824	&	0.152	&	103	&	7.372	&	0.084	&	91	&	1.122	&	0.102	&	103	&	5.903	&	0.145	&	103	&	5.101	&	0.090	\\
114	&	1.921	&	0.026	&	104	&	6.524	&	0.132	&	104	&	7.011	&	0.101	&	93	&	0.607	&	0.093	&	104	&	5.565	&	0.165	&	104	&	5.051	&	0.104	\\
115	&	2.097	&	0.029	&	107	&	5.275	&	0.152	&	108	&	7.377	&	0.093	&	94	&	1.390	&	0.107	&	108	&	6.176	&	0.220	&	108	&	5.534	&	0.097	\\
116	&	2.029	&	0.030	&	108	&	5.607	&	0.149	&	114	&	10.192	&	0.094	&	96	&	1.012	&	0.093	&	114	&	9.047	&	0.131	&	114	&	8.554	&	0.099	\\
117	&	2.623	&	0.036	&	109	&	5.336	&	0.144	&	115	&	15.155	&	0.141	&	97	&	0.782	&	0.102	&	115	&	13.890	&	0.168	&	115	&	13.512	&	0.144	\\
118	&	2.263	&	0.031	&	114	&	5.476	&	0.145	&	118	&	10.134	&	0.190	&	98	&	1.127	&	0.102	&	118	&	9.390	&	0.222	&	118	&	8.527	&	0.191	\\
119	&	2.060	&	0.028	&	115	&	4.956	&	0.142	&	134	&	7.277	&	0.078	&	99	&	1.866	&	0.097	&	134	&	6.963	&	0.191	&	134	&	5.592	&	0.081	\\
120	&	1.947	&	0.027	&	118	&	4.178	&	0.145	&	135	&	5.629	&	0.140	&	100	&	1.341	&	0.093	&	135	&	5.224	&	0.195	&	135	&	3.902	&	0.142	\\
121	&	2.073	&	0.029	&	134	&	5.900	&	0.147	&	136	&	5.200	&	0.127	&	101	&	1.634	&	0.098	&	136	&	4.208	&	0.175	&	136	&	3.440	&	0.129	\\
122	&	2.064	&	0.028	&	135	&	4.344	&	0.158	&	137	&	6.475	&	0.078	&	102	&	1.264	&	0.113	&	137	&	5.728	&	0.138	&	137	&	4.788	&	0.081	\\
123	&	2.147	&	0.034	&	136	&	4.806	&	0.156	&	138	&	7.119	&	0.075	&	103	&	1.469	&	0.119	&	138	&	5.995	&	0.131	&	138	&	5.443	&	0.079	\\

\hline
\end{tabular}
\label{tab:table-lcdata}
}
\end{table*}

\begin{table*}
\contcaption{}

\resizebox{19cm}{!}{
\begin{tabular}{cccccccccccccccccc} \hline

\multicolumn{3}{c}{$B$-band} &
\multicolumn{3}{c}{$R$-band} &
\multicolumn{3}{c|}{$S \, II$-band} &
\multicolumn{3}{c}{$\mathrm{H\alpha}$-band} &
\multicolumn{3}{c}{$\mathrm{H\alpha}$ line} &
\multicolumn{3}{c}{$\mathrm{{H\alpha}^\prime}$ line} \\

MJD & F  & $\sigma$ & MJD & F  & $\sigma$ &  MJD & F  & $\sigma$ & MJD & F  & $\sigma$ & MJD & F  & $\sigma$ & MJD & F  & $\sigma$  \\

\hline

124	&	2.578	&	0.041	&	137	&	5.051	&	0.153	&	140	&	11.560	&	0.127	&	104	&	1.446	&	0.130	&	140	&	10.761	&	0.173	&	140	&	9.855	&	0.129	\\
125	&	2.736	&	0.046	&	138	&	5.332	&	0.141	&	143	&	9.918	&	0.089	&	107	&	1.419	&	0.130	&	143	&	8.886	&	0.141	&	143	&	7.850	&	0.094	\\
126	&	3.430	&	0.076	&	140	&	5.659	&	0.142	&	148	&	12.372	&	0.115	&	108	&	1.200	&	0.199	&	$-$	&	$-$	&	$-$	&	148	&	9.862	&	0.121	\\
127	&	1.853	&	0.029	&	141	&	5.582	&	0.145	&	149	&	11.212	&	0.114	&	114	&	1.145	&	0.092	&	$-$	&	$-$	&	$-$	&	149	&	9.2	&	0.118	\\
128	&	1.814	&	0.028	&	142	&	4.859	&	0.145	&	152	&	8.924	&	0.089	&	115	&	1.265	&	0.090	&	$-$	&	$-$	&	$-$	&	152	&	6.968	&	0.092	\\
129	&	2.007	&	0.034	&	143	&	4.819	&	0.152	&	161	&	7.192	&	0.096	&	118	&	0.744	&	0.116	&	$-$	&	$-$	&	$-$	&	161	&	5.05	&	0.1	\\
130	&	1.773	&	0.029	&	148	&	4.507	&	0.142	&	162	&	7.353	&	0.115	&	119	&	1.184	&	0.199	&	$-$	&	$-$	&	$-$	&	162	&	5.128	&	0.121	\\
131	&	1.818	&	0.030	&	149	&	4.842	&	0.149	&	163	&	4.197	&	0.161	&	130	&	0.627	&	0.185	&	$-$	&	$-$	&	$-$	&	163	&	2.098	&	0.164	\\
132	&	1.993	&	0.039	&	150	&	6.303	&	0.160	&	164	&	7.345	&	0.099	&	132	&	0.778	&	0.123	&	$-$	&	$-$	&	$-$	&	164	&	5.336	&	0.105	\\
134	&	2.128	&	0.034	&	151	&	5.241	&	0.154	&	165	&	7.669	&	0.101	&	134	&	0.314	&	0.174	&	$-$	&	$-$	&	$-$	&	165	&	5.628	&	0.107	\\
135	&	1.891	&	0.037	&	152	&	5.352	&	0.148	&	166	&	8.542	&	0.089	&	135	&	0.404	&	0.135	&	$-$	&	$-$	&	$-$	&	166	&	6.752	&	0.094	\\
136	&	1.897	&	0.031	&	153	&	5.214	&	0.144	&	167	&	9.551	&	0.091	&	136	&	0.991	&	0.120	&	$-$	&	$-$	&	$-$	&	167	&	7.738	&	0.094	\\
137	&	1.856	&	0.027	&	161	&	5.056	&	0.141	&	168	&	10.889	&	0.101	&	137	&	0.747	&	0.114	&	$-$	&	$-$	&	$-$	&	168	&	9.175	&	0.103	\\
138	&	1.945	&	0.027	&	162	&	5.301	&	0.181	&	169	&	11.479	&	0.091	&	138	&	1.124	&	0.107	&	$-$	&	$-$	&	$-$	&	169	&	9.767	&	0.094	\\
139	&	1.993	&	0.027	&	163	&	6.693	&	0.148	&	170	&	8.754	&	0.091	&	140	&	0.798	&	0.117	&	$-$	&	$-$	&	$-$	&	170	&	6.848	&	0.095	\\
140	&	2.032	&	0.028	&	164	&	5.353	&	0.151	&	$-$	&	$-$	&	$-$	&	141	&	1.216	&	0.106	&	$-$	&	$-$	&	$-$	&	$-$	&	$-$	&	$-$	\\
141	&	1.690	&	0.023	&	165	&	4.774	&	0.175	&	$-$	&	$-$	&	$-$	&	142	&	1.227	&	0.092	&	$-$	&	$-$	&	$-$	&	$-$	&	$-$	&	$-$	\\
142	&	1.877	&	0.026	&	167	&	5.670	&	0.140	&	$-$	&	$-$	&	$-$	&	143	&	1.033	&	0.109	&	$-$	&	$-$	&	$-$	&	$-$	&	$-$	&	$-$	\\
143	&	1.948	&	0.026	&	168	&	4.985	&	0.149	&	$-$	&	$-$	&	$-$	&	$-$	&	$-$	&	$-$	&	$-$	&	$-$	&	$-$	&	$-$	&	$-$	&	$-$	\\
144	&	1.936	&	0.028	&	169	&	5.781	&	0.143	&	$-$	&	$-$	&	$-$	&	$-$	&	$-$	&	$-$	&	$-$	&	$-$	&	$-$	&	$-$	&	$-$	&	$-$	\\
145	&	1.969	&	0.025	&	170	&	4.531	&	0.150	&	$-$	&	$-$	&	$-$	&	$-$	&	$-$	&	$-$	&	$-$	&	$-$	&	$-$	&	$-$	&	$-$	&	$-$	\\
146	&	2.389	&	0.033	&	182	&	6.546	&	0.151	&	$-$	&	$-$	&	$-$	&	$-$	&	$-$	&	$-$	&	$-$	&	$-$	&	$-$	&	$-$	&	$-$	&	$-$	\\
147	&	2.898	&	0.043	&	$-$	&	$-$	&	$-$	&	$-$	&	$-$	&	$-$	&	$-$	&	$-$	&	$-$	&	$-$	&	$-$	&	$-$	&	$-$	&	$-$	&	$-$	\\
148	&	2.324	&	0.035	&	$-$	&	$-$	&	$-$	&	$-$	&	$-$	&	$-$	&	$-$	&	$-$	&	$-$	&	$-$	&	$-$	&	$-$	&	$-$	&	$-$	&	$-$	\\
149	&	2.259	&	0.029	&	$-$	&	$-$	&	$-$	&	$-$	&	$-$	&	$-$	&	$-$	&	$-$	&	$-$	&	$-$	&	$-$	&	$-$	&	$-$	&	$-$	&	$-$	\\
150	&	2.527	&	0.037	&	$-$	&	$-$	&	$-$	&	$-$	&	$-$	&	$-$	&	$-$	&	$-$	&	$-$	&	$-$	&	$-$	&	$-$	&	$-$	&	$-$	&	$-$	\\
151	&	2.469	&	0.029	&	$-$	&	$-$	&	$-$	&	$-$	&	$-$	&	$-$	&	$-$	&	$-$	&	$-$	&	$-$	&	$-$	&	$-$	&	$-$	&	$-$	&	$-$	\\
152	&	2.473	&	0.032	&	$-$	&	$-$	&	$-$	&	$-$	&	$-$	&	$-$	&	$-$	&	$-$	&	$-$	&	$-$	&	$-$	&	$-$	&	$-$	&	$-$	&	$-$	\\
153	&	2.576	&	0.031	&	$-$	&	$-$	&	$-$	&	$-$	&	$-$	&	$-$	&	$-$	&	$-$	&	$-$	&	$-$	&	$-$	&	$-$	&	$-$	&	$-$	&	$-$	\\
154	&	2.170	&	0.030	&	$-$	&	$-$	&	$-$	&	$-$	&	$-$	&	$-$	&	$-$	&	$-$	&	$-$	&	$-$	&	$-$	&	$-$	&	$-$	&	$-$	&	$-$	\\
155	&	2.292	&	0.031	&	$-$	&	$-$	&	$-$	&	$-$	&	$-$	&	$-$	&	$-$	&	$-$	&	$-$	&	$-$	&	$-$	&	$-$	&	$-$	&	$-$	&	$-$	\\
156	&	2.447	&	0.031	&	$-$	&	$-$	&	$-$	&	$-$	&	$-$	&	$-$	&	$-$	&	$-$	&	$-$	&	$-$	&	$-$	&	$-$	&	$-$	&	$-$	&	$-$	\\
157	&	2.111	&	0.027	&	$-$	&	$-$	&	$-$	&	$-$	&	$-$	&	$-$	&	$-$	&	$-$	&	$-$	&	$-$	&	$-$	&	$-$	&	$-$	&	$-$	&	$-$	\\
158	&	2.021	&	0.026	&	$-$	&	$-$	&	$-$	&	$-$	&	$-$	&	$-$	&	$-$	&	$-$	&	$-$	&	$-$	&	$-$	&	$-$	&	$-$	&	$-$	&	$-$	\\
159	&	2.136	&	0.029	&	$-$	&	$-$	&	$-$	&	$-$	&	$-$	&	$-$	&	$-$	&	$-$	&	$-$	&	$-$	&	$-$	&	$-$	&	$-$	&	$-$	&	$-$	\\
161	&	2.570	&	0.045	&	$-$	&	$-$	&	$-$	&	$-$	&	$-$	&	$-$	&	$-$	&	$-$	&	$-$	&	$-$	&	$-$	&	$-$	&	$-$	&	$-$	&	$-$	\\
162	&	2.424	&	0.036	&	$-$	&	$-$	&	$-$	&	$-$	&	$-$	&	$-$	&	$-$	&	$-$	&	$-$	&	$-$	&	$-$	&	$-$	&	$-$	&	$-$	&	$-$	\\
163	&	2.320	&	0.040	&	$-$	&	$-$	&	$-$	&	$-$	&	$-$	&	$-$	&	$-$	&	$-$	&	$-$	&	$-$	&	$-$	&	$-$	&	$-$	&	$-$	&	$-$	\\
164	&	2.358	&	0.041	&	$-$	&	$-$	&	$-$	&	$-$	&	$-$	&	$-$	&	$-$	&	$-$	&	$-$	&	$-$	&	$-$	&	$-$	&	$-$	&	$-$	&	$-$	\\
165	&	2.066	&	0.034	&	$-$	&	$-$	&	$-$	&	$-$	&	$-$	&	$-$	&	$-$	&	$-$	&	$-$	&	$-$	&	$-$	&	$-$	&	$-$	&	$-$	&	$-$	\\
167	&	2.094	&	0.026	&	$-$	&	$-$	&	$-$	&	$-$	&	$-$	&	$-$	&	$-$	&	$-$	&	$-$	&	$-$	&	$-$	&	$-$	&	$-$	&	$-$	&	$-$	\\
168	&	1.980	&	0.027	&	$-$	&	$-$	&	$-$	&	$-$	&	$-$	&	$-$	&	$-$	&	$-$	&	$-$	&	$-$	&	$-$	&	$-$	&	$-$	&	$-$	&	$-$	\\
169	&	1.977	&	0.025	&	$-$	&	$-$	&	$-$	&	$-$	&	$-$	&	$-$	&	$-$	&	$-$	&	$-$	&	$-$	&	$-$	&	$-$	&	$-$	&	$-$	&	$-$	\\
170	&	2.201	&	0.030	&	$-$	&	$-$	&	$-$	&	$-$	&	$-$	&	$-$	&	$-$	&	$-$	&	$-$	&	$-$	&	$-$	&	$-$	&	$-$	&	$-$	&	$-$	\\
171	&	2.164	&	0.027	&	$-$	&	$-$	&	$-$	&	$-$	&	$-$	&	$-$	&	$-$	&	$-$	&	$-$	&	$-$	&	$-$	&	$-$	&	$-$	&	$-$	&	$-$	\\
172	&	1.974	&	0.025	&	$-$	&	$-$	&	$-$	&	$-$	&	$-$	&	$-$	&	$-$	&	$-$	&	$-$	&	$-$	&	$-$	&	$-$	&	$-$	&	$-$	&	$-$	\\
173	&	1.828	&	0.025	&	$-$	&	$-$	&	$-$	&	$-$	&	$-$	&	$-$	&	$-$	&	$-$	&	$-$	&	$-$	&	$-$	&	$-$	&	$-$	&	$-$	&	$-$	\\

\hline
\end{tabular}
}
\end{table*}

\section{Analysis}\label{sec:analysis}

\subsection{Light curves}

\begin{figure*}
\resizebox{12cm}{15cm}{\includegraphics{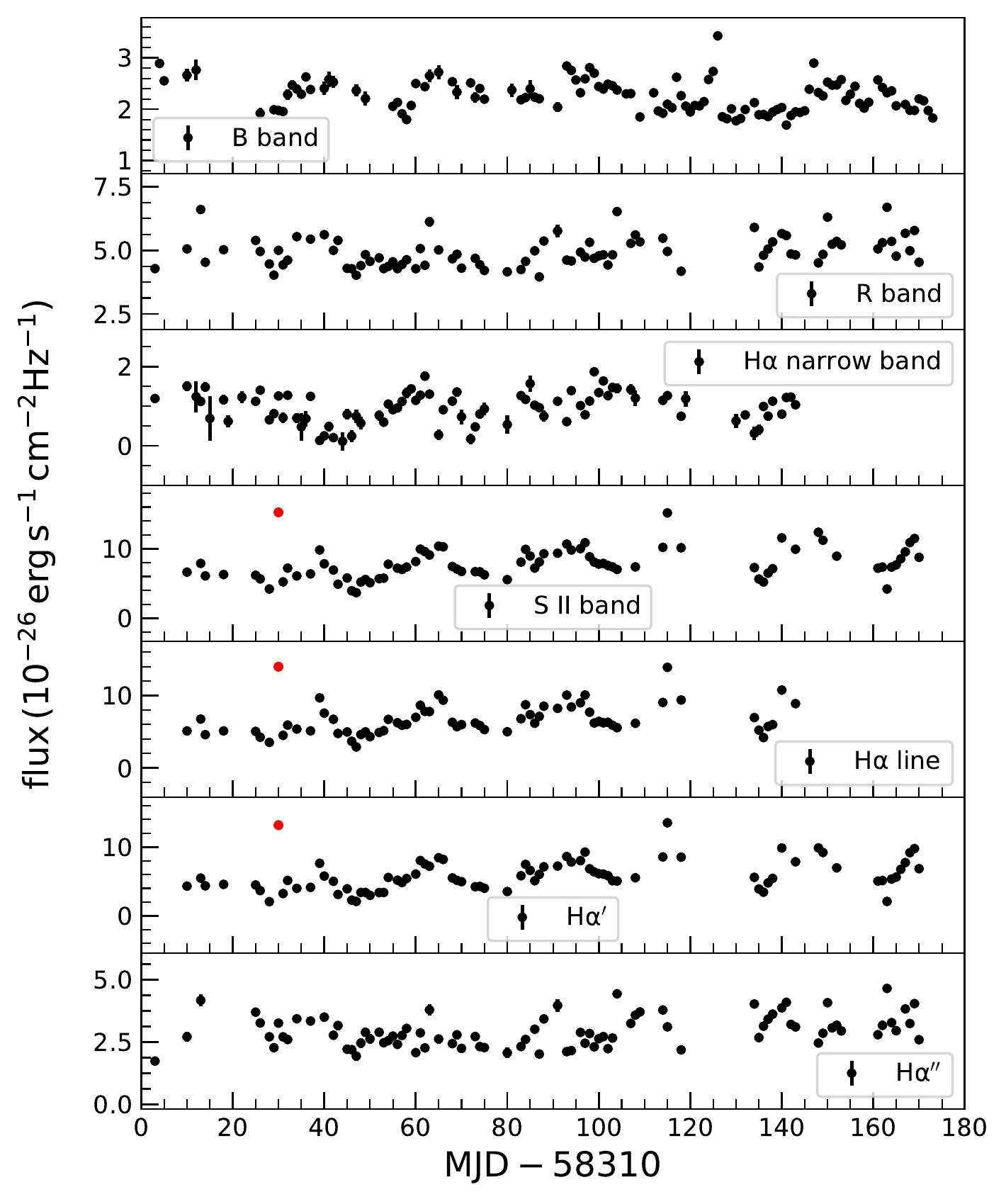}}
\caption{The light curves of Mrk 590 in the broad B- and R-band filters and narrow $\mathrm{H\alpha}-$ and S II band filters for the period from July, 2018 to December, 2018. The flux in the narrow $\mathrm{H\alpha}$-band was subtracted from the S II-band fluxes to get the $\mathrm{H\alpha}$ line light curve, whereas $\mathrm{H\alpha^{\prime}}$ and $\mathrm{H\alpha^{\prime\prime}}$ light curves were obtained considering a PL spectrum of the accretion disk and subtracted from the narrow S II and broad R-band, respectively.  The outlier is shown by red circle.}
\label{fig:fig-4}
\end{figure*}

The light curves of Mrk 590 in the optical $B$, $R$ and narrow $\mathrm{H\alpha}$ and S II bands and $\mathrm{H\alpha}$ line generated using the procedures from section \ref{sec:genlc} are shown in Fig. \ref{fig:fig-4}. The errors in the flux measurements are obtained through the standard propagation of errors. The top four panels in Fig. \ref{fig:fig-4} show the observed light curves in the broad $B$, $R$ and narrow $\mathrm{H\alpha}$ and S II filters, while the subsequent three panels show the $\mathrm{H\alpha}$ line light curves derived using Equation \ref{eq:hlp_ori}, \ref{eq:cont_pow} and \ref{eq:hlp_R}, respectively. The $\mathrm{H\alpha}$ line light curve obtained from the broad $R$-band data is noisy, which indicates the difficulty in getting line fluxes from broad band observations. There is negligible contamination from the radio jet to the observed flux variations as the source is a Seyfert galaxy and considered to be radio-quiet. The characteristics of the light curves are given in Table \ref{tab:table-var}. To characterise the variability, we used the normalized excess variance  \citep[$F_{\mathrm{var}}$;][]{2002ApJ...568..610E, 2003MNRAS.345.1271V, 2017MNRAS.466.3309R}. The uncertainties in $F_{\mathrm{var}}$ were obtained following \citet{2002ApJ...568..610E}. $R_{\mathrm{max}}$ and $\sigma$ represent the ratio between the maximum and minimum flux and standard deviation of the light curves, respectively. We calculated the host galaxy ($L_{\mathrm{host}}$) and AGN luminosity ($L_{\mathrm{AGN}}$) from the average fluxes obtained for the host galaxy and AGN, respectively, for all the bands and $\mathrm{H\alpha}$ line.

\begin{table*}
\caption{Variability statistics in $B$, $R$, narrow $\mathrm{H\alpha}$ and S II bands and for $H_{\alpha}$ line  in observer's frame, where
$\mathrm{\lambda_{eff}}$ is the effective wavelength in Angstroms. The average values ($\langle f \rangle$)
and the standard deviation $\sigma$ of the light curves are in units of
$\mathrm{10^{-26} \, erg \, s^{-1} \, cm^{-2} \, Hz^{-1}}$. The luminosities in each band are in erg s$^{-1}$.}
\label{tab:table-var}
\small
\begin{tabular}{lccccccl} \hline
Filter & $\lambda_{\mathrm{eff}}$ & $\langle f \rangle$ & $\sigma$ & $F_{\mathrm{var}}$ & $R_{\mathrm{max}}$ & $\log \, L_{\mathrm{host}}$ & $\log \, L_{\mathrm{AGN}}$ \\ \hline
$B$ & 4363 & $2.26 \pm 0.05$ & 0.31 & $ 0.132 \pm  0.001$ & 2.03 & $42.22 \pm 0.01$ & $43.36 \pm 0.01$   \\

$R$ & 6410 & $4.95 \pm 0.15$ & 0.62 & $ 0.121 \pm  0.001$ & 1.69 & $43.98 \pm 0.01$ & $43.53 \pm 0.01$   \\

$\mathrm{H\alpha}$  & 6563 & $ 0.97 \pm 0.13$ & 0.40 & $0.379 \pm  0.005$ & 16.60 & $ 43.85 \pm  0.01 $ & $ 42.81 \pm  0.06 $  \\

S II & 6719 & $7.88 \pm 0.12$ & 2.26 & $0.286 \pm 0.001$ & 4.17 & $43.70 \pm 0.01$ & $43.71 \pm 0.01$ \\

$\mathrm{H\alpha}$ line & 6564.6 & $6.73 \pm 0.17$ & 2.17 & $0.321 \pm 0.001$ & 4.80 & $-$ & $43.65 \pm 0.01$  \\

$\mathrm{{H\alpha}^{\prime}}$ line & 6564.6 & $5.90 \pm 0.13$ & 2.28 & $0.385 \pm 0.001$ & 6.56 & $-$ & $43.60 \pm 0.01$  \\

$\mathrm{{H\alpha}^{\prime\prime}}$ line & 6564.6 & $2.94 \pm 0.16$ & 0.63 & $0.206 \pm 0.001$ & 2.68 & $-$ & $43.29 \pm 0.02$  \\
\hline

\end{tabular}

\end{table*}

\subsection{Cross-correlation function analysis}

\subsubsection{B-band v/s narrow band light curves}

\begin{figure}
\resizebox{9cm}{7cm}{\includegraphics{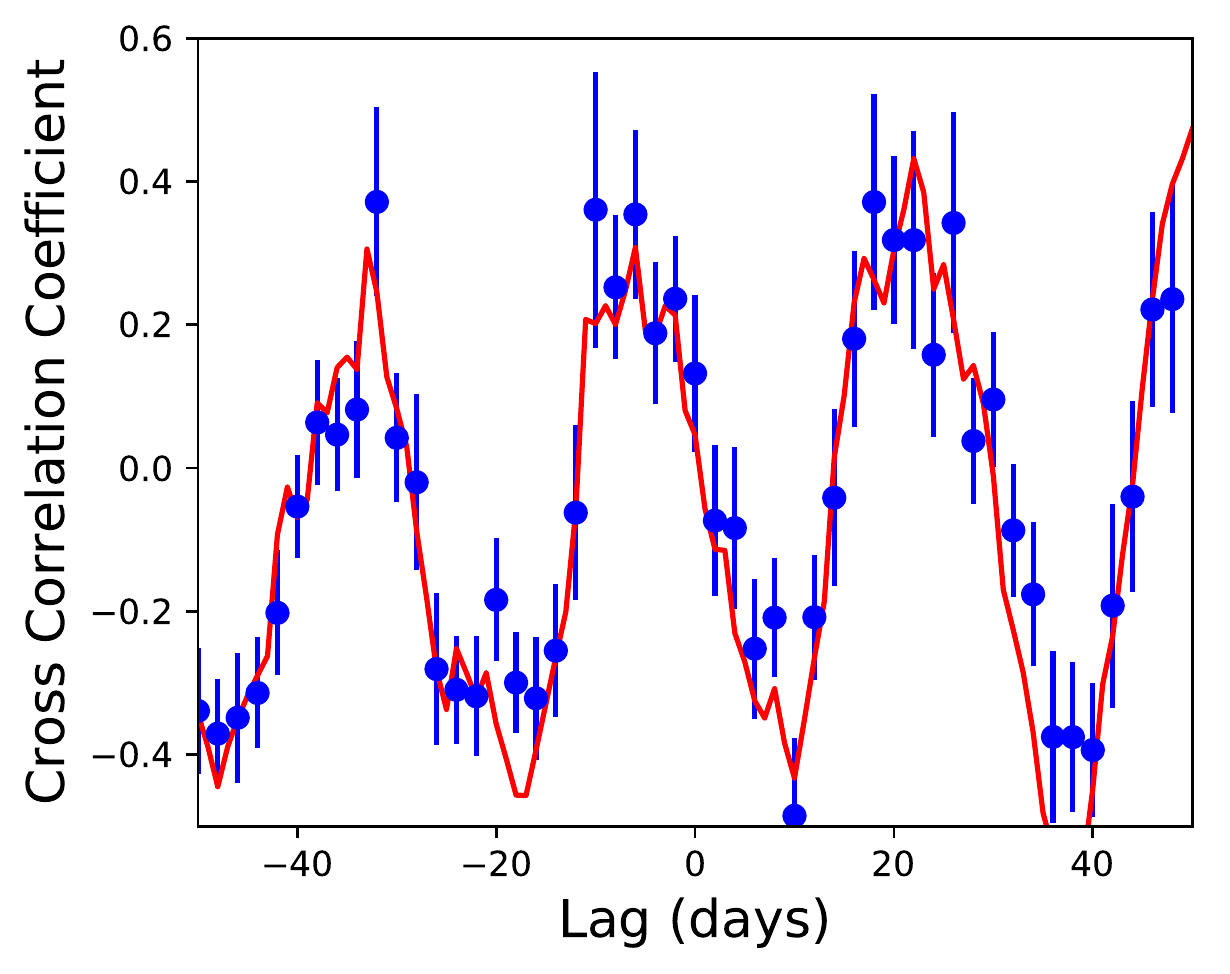}}
\caption{The CCFs between B and  $\mathrm{H\alpha}$ light curves. The solid red line represents the ICCF, points with error bars show the DCF obtained using $\Delta\tau = 1$ day.}
\label{fig:fig-ccf_range}
\end{figure}

\begin{figure}

\includegraphics[scale=0.6]{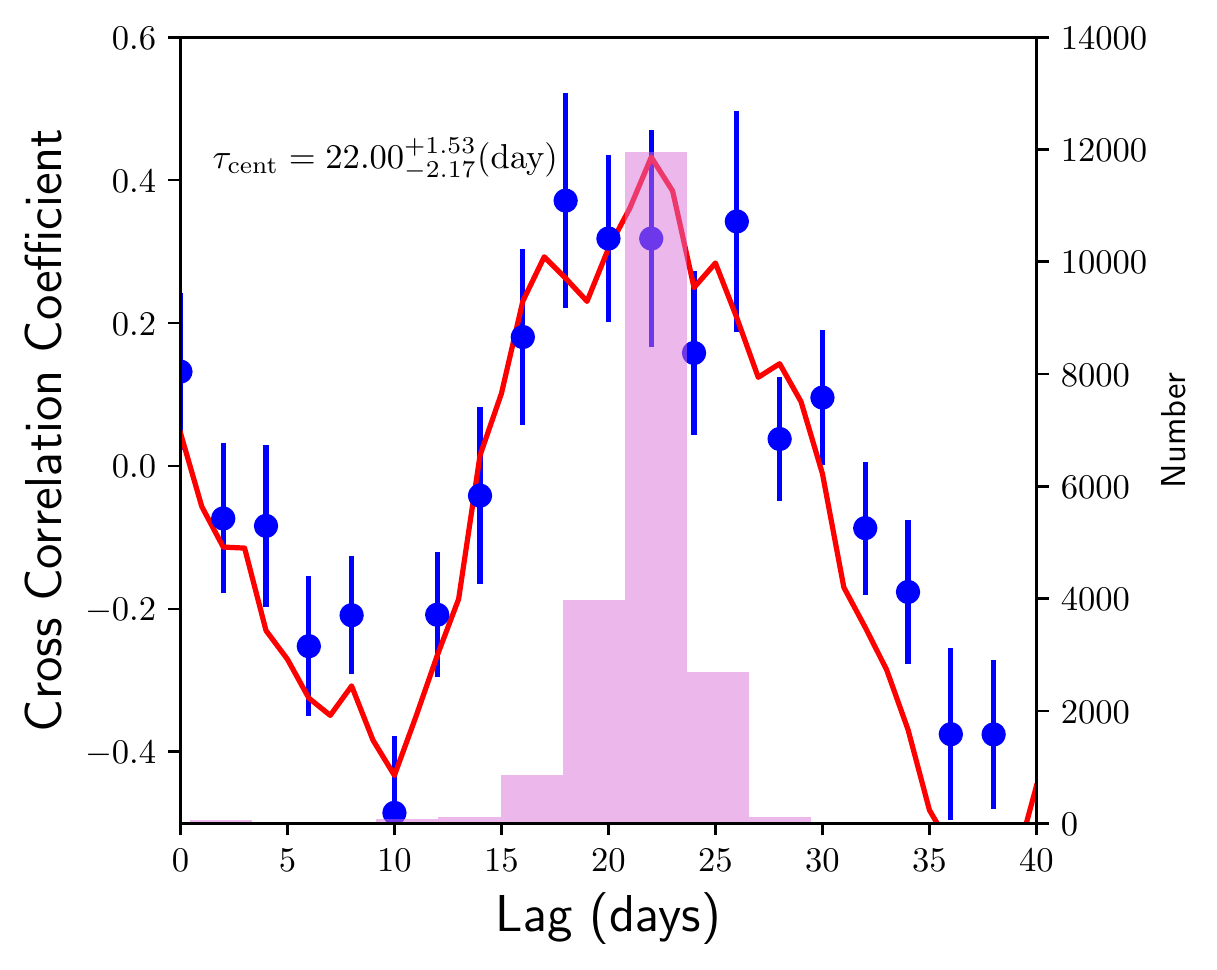}
\includegraphics[scale=0.6]{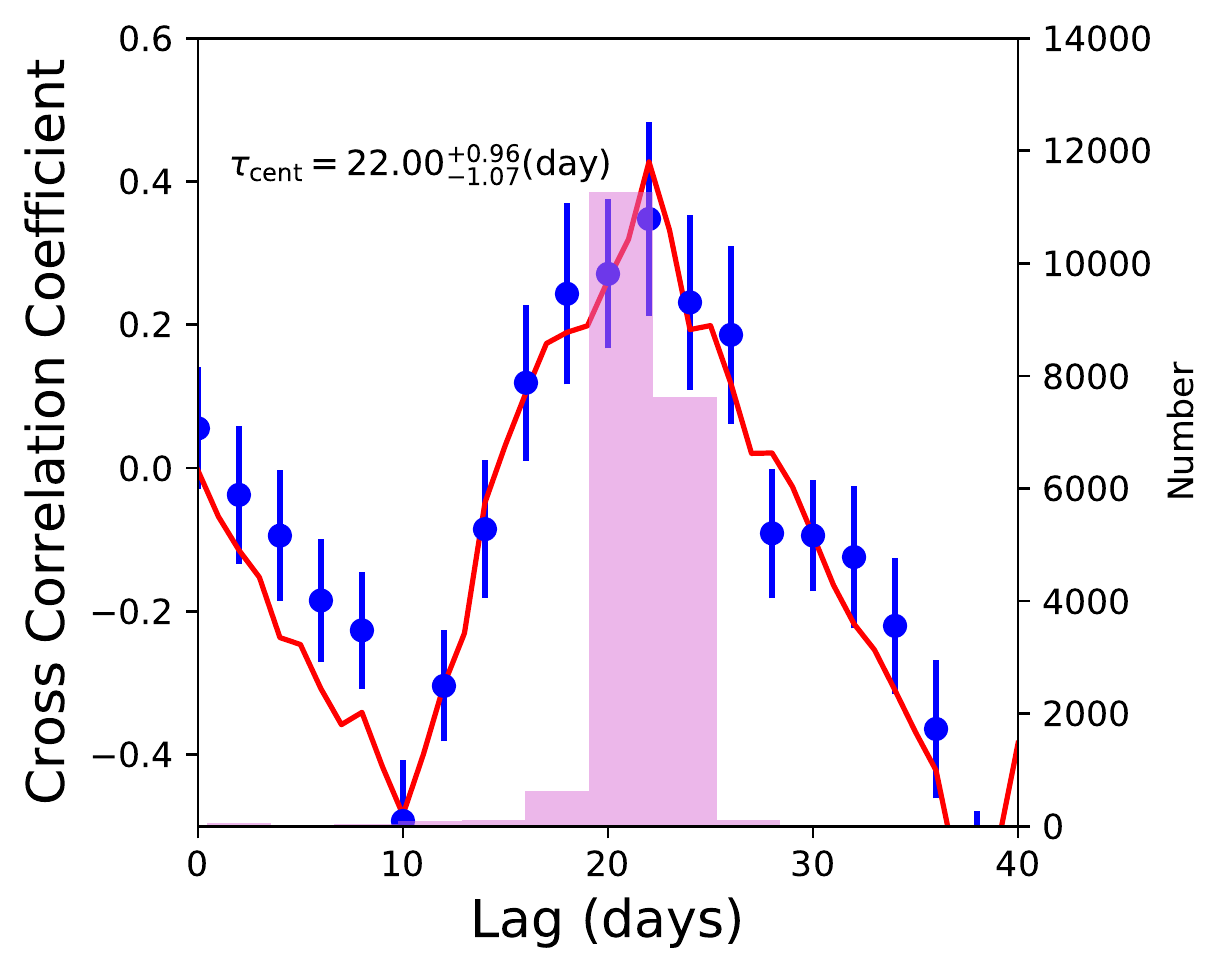}

\caption{The CCFs between $B$ and $\mathrm{H\alpha}$ line flux (top panel) and the same with $\alpha=1/3$ (bottom panel). The solid red line represents the ICCF, points with error bars show the DCF obtained using $\Delta\tau = 1$ day
and the corresponding distribution of the $\tau_{cent}$ obtained using 20000 Monte Carlo simulations is shown by histogram in each panel. The value of $\tau_{cent}$ from the median of the distribution obtained using Monte Carlo simulation using ICCF is noted in each panel.}
\label{fig:fig-ccf}
\end{figure}

Mrk 590 shows strong variations in the optical $B$-band and H$\alpha$ line light curves (see Fig. \ref{fig:fig-4}). The $\mathrm{H\alpha}$ light curves used here were derived (a) by subtracting
the observed fluxes in the narrow $\mathrm{H\alpha}$ band from that measured in the narrow S II band and (b) by subtracting the continuum obtained by extrapolating the observed B-band flux using $\alpha = 1/3$ from the measured S II fluxes. To calculate the time lag between the $B$-band and the H$\alpha$ line light curves, we employed two well-known methods, namely the interpolated cross-correlation function \citep[ICCF;][]{1986ApJ...305..175G,1987ApJS...65....1G} and the discrete cross-correlation function \citep[DCF;][]{1988ApJ...333..646E}. The total duration of S II-band data is $\sim$160 days. Hence, we  first evaluated the CCF in the range $-50$ days to $+50$ days considering the total duration of observation should exceed at least 3 times the expected maximum lag to locate the peak of the CCF \citep{2004AN....325..248P}. The CCFs (Fig. \ref{fig:fig-ccf_range}) show two peaks; one is at around $-$6 days and another is at around $+22$ days. Since a  negative lag is physically not possible, we used the centroid of the CCF around the positive peak at +22 days to estimate the expected BLR lag between the $B$-band and the $\mathrm{H\alpha}$ line. The centroid position was estimated by considering the points that are within 80 per cent of the maximum of the CCF as shown in Fig. \ref{fig:fig-ccf}. The cross-correlation centroid distribution (CCCD) obtained using ICCF via Monte Carlo simulation based on the flux randomization and random subset selection  described in \citet{1998PASP..110..660P} for 20000 iterations is shown in the histogram in each panel of Fig. \ref{fig:fig-ccf}. The median of the CCCD is taken as a representation of the lag, which is found to be about 22.00 days in the observed frame. Since the distribution has a non-Gaussian shape, we calculated the uncertainties within 68\% confidence interval around the median value as described in \citet{2018MNRAS.475.5330M}.

\begin{figure}

\includegraphics[scale=0.6]{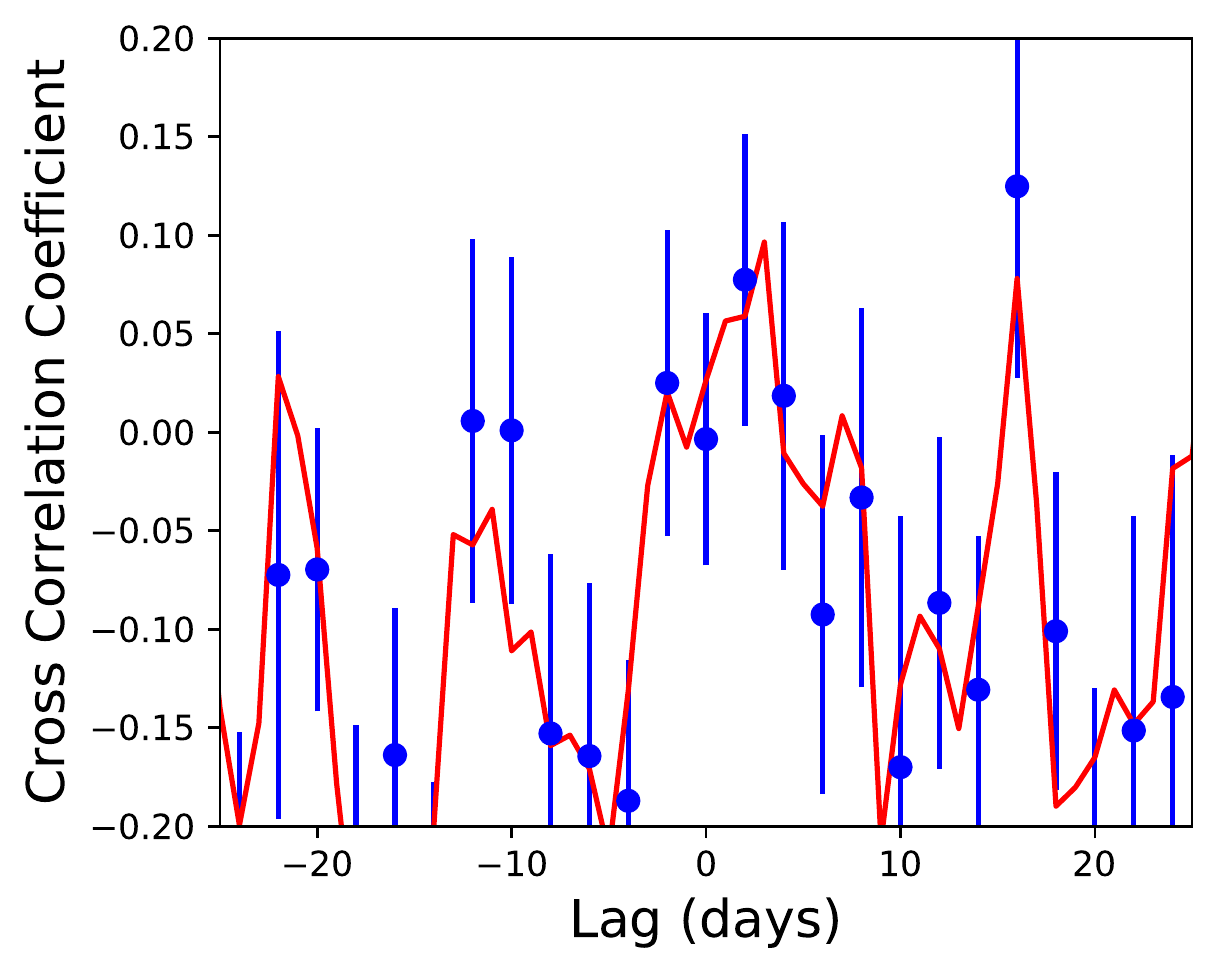}
\includegraphics[scale=0.6]{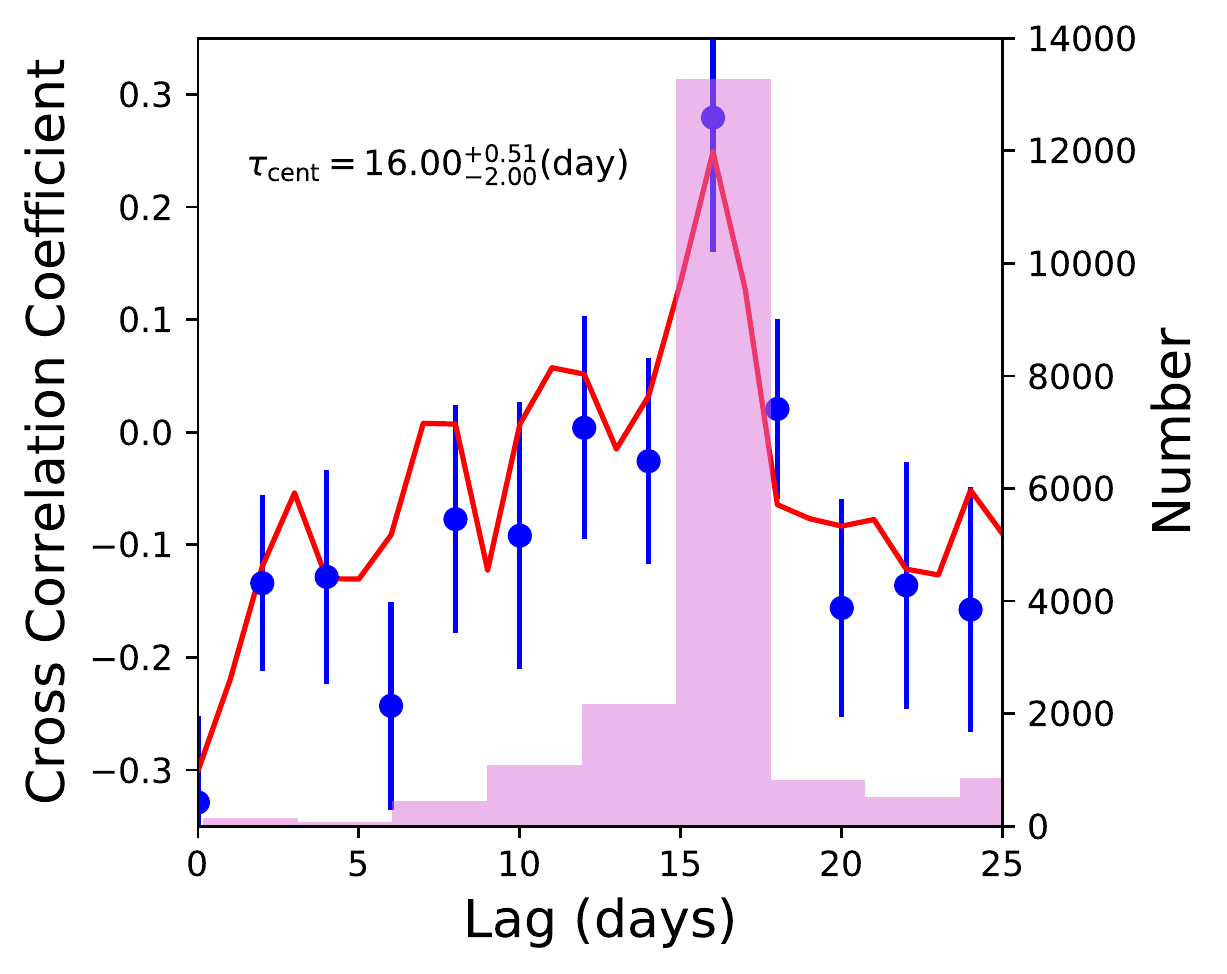}

\caption{The CCFs between $B$ and $R$ (top panel) and continuum subtracted $R$ ($\mathrm{H\alpha}^{\prime\prime}$) with $\alpha=1/3$  (bottom panel) are shown. The solid red line represents the ICCF, points with error bars show the DCF obtained using $\Delta\tau = 1$ day and the corresponding distribution of the $\tau_{\mathrm{cent}}$ obtained using 20000 Monte Carlo Simulations is shown by histogram. The value of $\tau_{\mathrm{cent}}$ from the median of the distribution obtained using Monte Carlo simulation using ICCF is also noted in the bottom panel.}
\label{fig:fig-ccf_R}
\end{figure}

The measured lags and their uncertainties between the $B$-band and $\mathrm{H\alpha}$ line light curves in the observed frame derived from the Monte Carlo simulation using the ICCF analysis method are given in Table \ref{tab:table-ccfres}. We found similar time lags between the $B$-band continuum and the $\mathrm{H\alpha}$ light curves derived from two approaches. Considering the $\mathrm{H\alpha}$ line light curve derived using
Equation \ref{eq:hlp_ori}, which is straight forward devoid of any uncertainties due to the use of a single value of $\alpha$ as in Equations \ref{eq:cont_pow} and \ref{eq:hlp_R}, we found the rest frame time lag between the $B$-band and the $\mathrm{H\alpha}$ line as $21.44^{+1.49}_{-2.11}$ days, which corresponds to a BLR size of 0.018 pc. We noticed one outlier point (shown as red circle) in the S II, $\mathrm{H\alpha}$ and $\mathrm{H\alpha^{\prime}}$ light curves in Fig. \ref{fig:fig-4}. On examining the S II band image acquired at that epoch, we found the target source to be affected by a weak
cosmic ray hit. After excluding this outlier point in the S II, $\mathrm{H\alpha}$, and $\mathrm{H\alpha^{\prime}}$ light curves, we carried out the CCF analysis. We found a rest-frame lag of $21.42^{+1.03}_{-2.38}$ days, which is similar to that
obtained with the same outlier point included. The impact of this sole outlier point is therefore, minimal in our measured lag.

\subsubsection{$B$-band v/s $R$-band light curves}

The CCF between the $B$ and $R$-band light curves is shown in Fig. \ref{fig:fig-ccf_R} (top). The host corrected $R$-band has contributions from the continuum as well as the $\mathrm{H\alpha}$ line fluxes (see Fig. \ref{fig:spec}). This is also reflected in the CCF (Fig. \ref{fig:fig-ccf_R}; top panel), which shows two prominent peaks; one at a smaller lag of $\sim$3 days due to the contribution from the continuum flux originating from the accretion disk, and another at lag of $\sim$16 days, which is due to the contribution from the H$\alpha$ line flux from the BLR. Recent disk reverberation mapping observations utilizing high quality data on Fairall 9 and Mrk 110 by \citet{2020MNRAS.498.5399H} and \citet{2021MNRAS.504.4337V}, respectively, show well correlated variations between different optical bands. However, we note a weak correlation between the continuum B and R-band light curves with a peak of CCF of $\mathrm{r_{max}} \sim$0.1. Such a low correlation is due to a combination of the following factors (a) our light curves are of moderate quality, (b) the B-band light curve is dominated by the AGN, while the R-band light curve is contaminated to a large extent by light from the host galaxy. Though we carried out \textsc{galfit}  to remove the contribution of the host galaxy to the observed R-band brightness, it was not effective as the seeing was not stable during the observing period. As pointed by \citet{2000AJ....119.1534C} seeing fluctuations during the observations can severely affect the observed light curves. According to them, degradation in the light curves due to seeing fluctuations are larger for AGN with brighter host. In our case too, (i) compared to B-band, R-band has larger host contamination and (ii) seeing was not stable and has varied between 1.7 to 4.4 arcsec during the monitoring period. Such seeing variations coupled with the larger host galaxy contribution to the R-band measurements have led to poor quality R-band light curve. The B-band light curve when cross-correlated with the poor quality R-band light curve has led to the weak correlation. The R-band light curve having little resemblance to the B-band light curve is also evident in Fig. \ref{fig:fig-4} and (c) the R-band contains contributions from two spatially different regions, namely, the continuum coming from the accretion disk and the redshifted $\mathrm{H\alpha}$ from the BLR.

To have only the $\mathrm{H\alpha}$ line contributing to the measured flux in the $R$-band, we subtracted the contribution of the continuum to the observed $R$-band flux using Equation \ref{eq:hlp_R}. We then, cross-correlated the $B$-band and continuum subtracted $R$-band ($\mathrm{H\alpha^{\prime\prime}}$) light curves. We found a lag of $\sim$16 days in the observed frame with an improved correlation of $\mathrm{r_{max}}$ $\sim$ 0.27 (Fig. \ref{fig:fig-ccf_R}; bottom panel). Such a weak correlation has also been found in other studies where reliable lag measurements have been claimed, e.g., broad band photometric RM has been carried out for NGC 4395 by \citet{2012ApJ...756...73E} between SDSS $g$, $r$ and $i$ filters, which provided a BLR size but with a weak correlation of $\mathrm{r_{max}}$ $\sim$ 0.1.  The measured lag with uncertainties between $B$ and continuum subtracted $R$-band is given in Table \ref{tab:table-ccfres}. The lag between $B$ and $\mathrm{H\alpha}$ line light curve obtained from continuum subtracted narrow S II band data is slightly larger compared to that of using continuum subtracted $R$-band as $\mathrm{H\alpha}$ line flux. This is because the narrow S II band has much smaller contribution from AGN continuum than in broad $R$-band \citep{2014ApJ...788..159K}.

 Photometric continuum RM observations \citep{2016ApJ...821...56F, 2018ApJ...854..107F, 2021arXiv210502884H} suggest that continuum lags ($\tau$) increase with wavelength ($\lambda$) with $\mathrm{\tau \, \propto \, \lambda^{4/3}}$  as expected from the standard accretion disk model by \citet{1973A&A....24..337S}.
Following \citet{2021arXiv210502884H}, the accretion disk size can be parameterized as: 
  
\begin{equation}
\tau_{optical} - \tau_{uv} = \tau_{0}  \big[\big(\frac{\lambda_{optical}}{2700 \, \text{\r{A}}}\big)^{\beta} - \big(\frac{\lambda_{uv}}{2700 \, \text{\r{A}}}\big)^{\beta}\big]
\label{eq:acn1}
\end{equation}  

where, $\tau$ represents the rest frame lag, $\mathrm{\lambda_{uv}}$ and $\mathrm{\lambda_{optical}}$ are the rest frame
UV and optical reference wavelengths, respectively, with $\beta = 4/3$ for a standard accretion disk \citep{1973A&A....24..337S}. Therefore, using Equation \ref{eq:acn1}, the lag between B and R-band continua can be obtained as:

\begin{equation}
\tau_{R} - \tau_{B} = \tau_{0}  \big[\big(\frac{\lambda_{R}}{2700 \, \text{\r{A}}}\big)^{\beta} - \big(\frac{\lambda_{B}}{2700 \, \text{\r{A}}}\big)^{\beta}\big]
\label{eq:acn2}
\end{equation} 

with the disk normalization parameter $\tau_{0}$ given as

\begin{equation}
\begin{aligned}
\tau_{0} = \frac{1}{c} \big(\frac{45G}{16 \pi^6 h c^2}\big)^{1/3} (2700 \, \text{\r{A}})^{4/3} \chi^{4/3} \big(\frac{A_{bol}}{\eta c^2}\big)^{1/3} \\ \langle M_{BH} \rangle^{1/3}  \langle\lambda L_{\lambda}\rangle^{1/3}
\label{eq:acn3}
\end{aligned}
\end{equation}

Here, we assumed the geometrical factor related to the flux-weighted mean radius, $\chi$ = 2.49 as suggested by \citet{2021arXiv210502884H}, radiative efficiency $\eta = 0.1$,  bolometric luminosity $\mathrm{L_{bol}}$ = $\mathrm{A_{bol} \lambda L_{\lambda}}$, where $\lambda L_{\lambda}$ was obtained from the monochromatic luminosity at 5100 \AA \ with $\mathrm{A_{bol}} = 8.1 \pm 0.4$ \citep{2012MNRAS.422..478R}. Using Equations \ref{eq:acn2} and \ref{eq:acn3}, we found the expected lag between B and R-band continua as $0.32^{+0.01}_{-0.01}$ days, where the uncertainties were obtained using the propagation of error method. The expected continuum lag from the continuum-lag$-$ wavelength relation is found to be very small compared to the BLR size obtained based on lag between B-band and continuum subtracted $\mathrm{H\alpha}$ line light curves. This implies that the inherent lag between continuum variations in B and R-band light curves has negligible effect on the BLR size deduced from cross-correlation between the B band continuum and the $\mathrm{H\alpha}$ line light curves.

\subsection{JAVELIN}
\begin{figure*}
\resizebox{16cm}{8cm}{\includegraphics{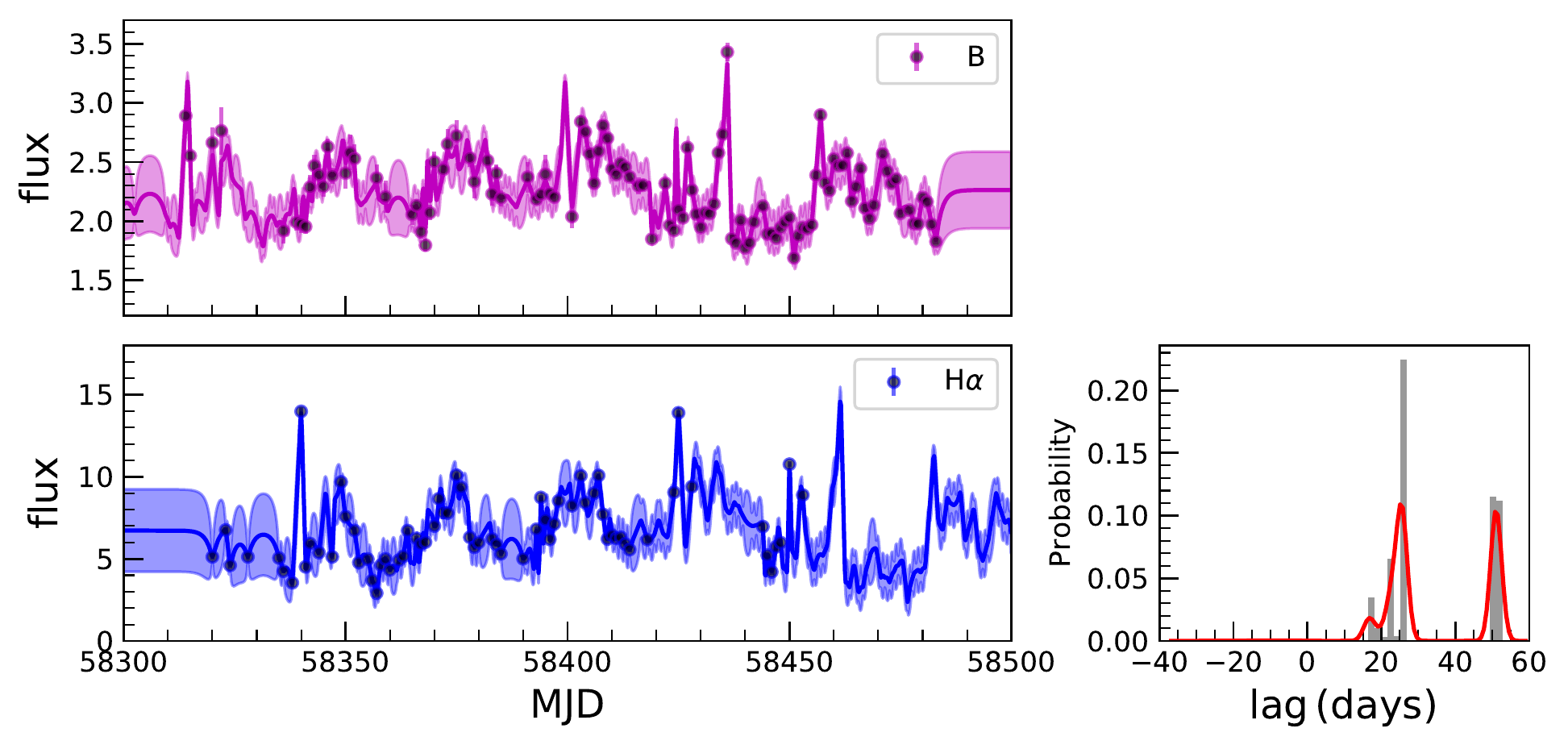}}
\caption{\textsc{javelin} fit to the B and H$\alpha$ light curves. The points with errorbars are the observed data while the solid line represents the best fit and shaded region shows 1$\sigma$ error. The lag probability distribution along with smooth kernel density are shown in the lower-right panel.}
\label{fig:javelin_fit}
\end{figure*}

Apart from CCF analysis, we also used the \textsc{javelin} code developed by \citet{2011ApJ...735...80Z} to calculate the time lag between $B$ and H$\alpha$ as well as the $B$-band and continuum subtracted $R$-band. In \textsc{javelin}, the driving continuum light curve is modeled using a damped random walk (DRW) process, which provides a satisfactory statistical description of the quasar variability \citep[e.g.,][]{2009ApJ...698..895K}. The DRW model is defined by an exponential covariance matrix with two parameters; amplitude and time scale of the variability. First, a top-hat response function representing the BLR response is convolved with the driving continuum resulting in an emission line light curve, which is a shifted, scaled and smoothed version of the driving continuum light curve. The probability distribution of model parameters and the best-fit model are found by maximizing the likelihood function through a Markov Chain Monte Carlo approach. Using  $B$-band as the driving continuum light curve, we searched for a lag within a range of $-40$ to 60 days. An example of the best-fit model light curve and the distribution of the lag obtained from fitting the $B$-band and H$\alpha$ line light curves using Equation \ref{eq:hlp_ori} are shown in Fig. \ref{fig:javelin_fit}. The lag distribution shows two peaks; the most prominent one is at $~$20 days and a secondary peak is at $~$50 days. Considering that the most prominent peak is at $~$20 days together with the result obtained from CCF analysis, we obtained a lag of $22.48 ^{+0.09}_{-0.89}$ days in the observed frame. We summarize the \textsc{javelin} lag results in Table \ref{tab:table-ccfres}. We found a lag of $\sim16$ days between $B$-band and continuum subtracted $R$-band light curves. 

\begin{figure*}
\resizebox{16cm}{8cm}{\includegraphics{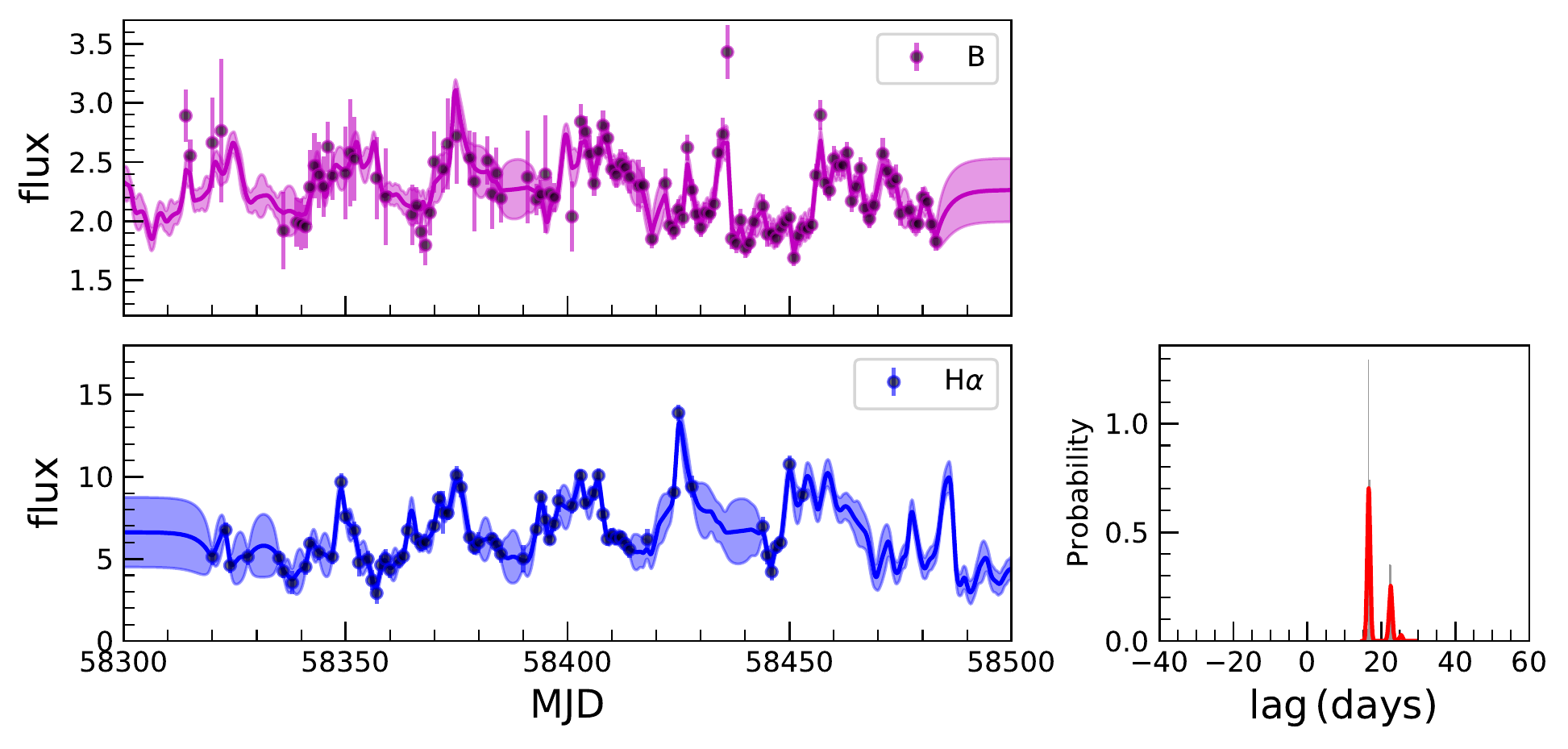}}
\caption{\textsc{javelin}  fit to the B and H$\alpha$ light curves after removing the outlier at MJD 583140 and inflating the errorbars
by a factor of 3.}
\label{fig:javelin_fit2}
\end{figure*}

 We note that \textsc{javelin} can over-fit the noise in the data if the error bars are too small. In fact,  \textsc{javelin} fit shows spurious peak in the B-band model light curve at MJD$~$58400 (see Fig. \ref{fig:javelin_fit}) having no observed data at that epoch. Therefore, we did two tests to check the validity of our \textsc{javelin} results; (1) we removed the  outlier at MJD 583140 from H$\alpha$ light curve, and performed the \textsc{javelin} fit, and obtained a lag of $22.52^{+0.37}_{-6.01}$ days in the observed frame, (2) after removing the outlier we also inflated the errorbars by a factor of 3. The light curves are well fitted with \textsc{javelin} providing a lag of $16.66^{+5.85}_{-0.19}$ days (see Fig. \ref{fig:javelin_fit2}) in the observed frame. Both the tests provide a consistent lag within the errorbars, and agree well with the one found from the original light curves. Therefore, the lag measurement using \textsc{javelin} is reliable. The lags obtained by \textsc{javelin} and ICCF are consistent with each other, therefore, in all further discussion, we consider the lags obtained from ICCF analysis.



\begin{table}
\caption{Lag measurements. Columns are (1) method used for continuum subtraction (2) filters (3-4)  $\tau_{cent}$ with errors in days as obtained by ICCF and JAVELIN in observer's frame. R-band fluxes were corrected for the continuum using power-law with slope $\alpha$.}
\label{tab:table-ccfres}
\small
\begin{tabular}{llll} \hline
method & filter &  ICCF & JAVELIN \\
(1)    &  (2)   &  (3)  &  (4) \\\hline
direct & B$-\mathrm{H\alpha}$  &$22.00^{+1.53}_{-2.17}$  & $25.42^{+0.11}_{-3.03}$ \\
PL subtracted ($\alpha = 1/3$)  & B$-\mathrm{H\alpha}$  & $22.00^{+0.96}_{-1.07}$ & $22.48^{+0.09}_{- 0.89}$  \\
PL subtracted ($\alpha = 1/3$) & B$-$R & $16.00^{+0.51}_{-2.00}$ & $16.41^{+0.09}_{-0.83}$\\

\hline

\end{tabular}

\end{table}

\section{DISCUSSION}\label{sec:discussion}

From the single-epoch spectrum of Mrk 590 we estimated the monochromatic continuum luminosity at 5100\AA \,  ($L_{5100}$) of $43.12\pm0.01$ erg s$^{-1}$ (see section \ref{sec:spectrum}). This corresponds to an expected BLR size of $\sim$12 days based on the best-fit size-luminosity relation of \citet{2013ApJ...767..149B}, which is a factor of 2 smaller than our measurement. We note that the $L_{5100}$ could be affected by some host galaxy contribution. Although the spectrum does not allow us to clearly separate the host galaxy contribution, photometric image decomposition method provides host galaxy contribution to the total flux, which is $\sim50$\% in the observed S II band data. The host galaxy corrected AGN luminosity in $B$-band is found to be $\log L_{\mathrm{AGN}} = 43.36 \pm 0.01$ erg s$^{-1}$,  which is in excellent agreement with the AGN luminosity of $\log L_{\mathrm{AGN}} = 43.31 \pm 0.05$ erg s$^{-1}$ at 5100 {\AA} wavelength recently obtained by \citet{2020MNRAS.491.4615K}.

\begin{figure}
\resizebox{8cm}{7cm}{\includegraphics{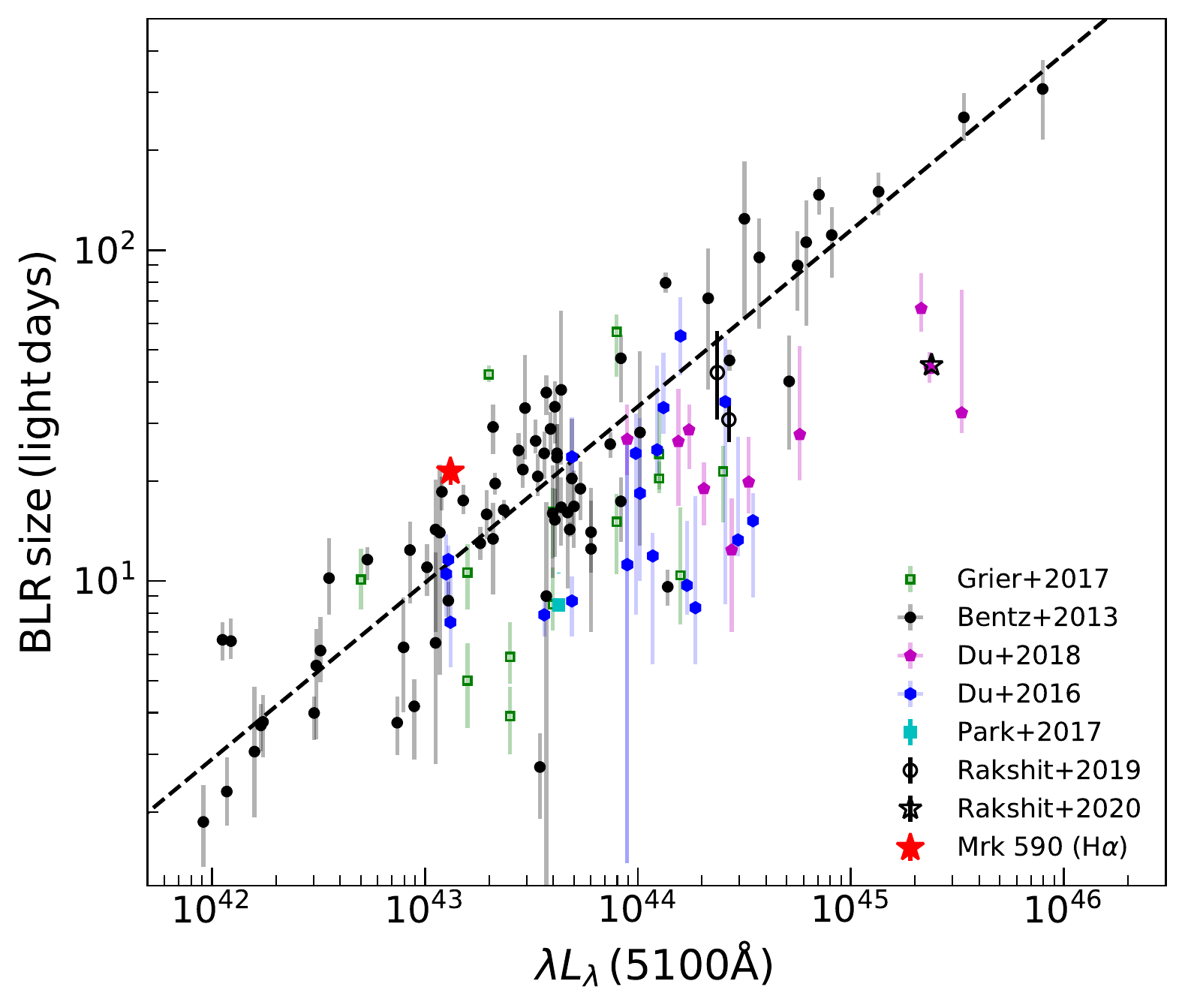}}
\caption{The BLR size-luminosity relation of Mrk 590 along with other AGN from literature \citep{2013ApJ...767..149B,2016ApJ...825..126D,2017ApJ...851...21G,2018ApJ...856....6D,2018ApJ...865....5R,2020arXiv200707672R}. The best-fit size-luminosity relation from \citet{2013ApJ...767..149B} is shown as dotted line.}
\label{fig:lag_lum}
\end{figure}

The position of Mrk 590 in the size-luminosity diagram is shown in Fig. \ref{fig:lag_lum} based on our H$\alpha$ BLR size measurement and $L_{5100}$ (host uncorrected) along with other AGN from the literature. Our measurement does not significantly deviate from the best-fit relation. \citet{1998ApJ...501...82P} performed H$\beta$ spectroscopic RM of Mrk 590 and estimated a range of BLR size depending on the observational epoch ranging from $14.0^{+8.5}_{-8.5}$ light-days  (MJD = 48848$-$49048) to $29.2^{+4.9}_{-5.0}$ light-days (MJD = 49183$-$49338). Therefore, our measured lag is consistent with the measurement of \cite{1998ApJ...501...82P}. Note that, our result is based on the H$\alpha$ lag. \cite{Bentz_2010} found that on-average H$\alpha$ lag is 1.54 times larger than the H$\beta$ lag. Therefore, the position of Mrk 590 in Fig. \ref{fig:lag_lum} is consistent with that expectation.

Our measured BLR size together with other measurements on Mrk 590 are found to be comparable with the inner radius of the dust torus found by \citet{2020MNRAS.491.4615K}. This could be due to two reasons; (a) the DRM observations were carried out during the period 2003 $-$ 2007 when the source was in a faint state. As a result, the drastic decrease in the ionizing continuum radiation from the accretion disk causes the dust particles to exist closer to the BLR region making the torus size comparable to the BLR size. According to \citet{2020MNRAS.491.4615K}, after the drastic drop in accretion disk luminosity, the dust replenishment is attained by formation of new dust grains in the BLR/innermost dust torus region, and (b) the distribution of dust particles in the torus region is not uniform as shown in the Unification scheme of AGN and may change over time. Because of the presence of a clumpy dust torus \citep{2013arXiv1301.4244S}, the obscuration of ionizing continuum by the torus will also vary over time which may cause the torus to lie close to the BLR over a certain period of time. The clumpy dust torus can even play an important role in making an AGN a "changing look AGN" due to the varying obscuration of the emission line fluxes from the BLR along the line of sight over time.

The spectra of Mrk 590 obtained by \cite{1998ApJ...501...82P} show clear presence of a broad H$\beta$ line having an FWHM of $\sim2300$ km s$^{-1}$. However, between 2006 $-$ 2013, its continuum luminosity decreased by a factor of 100 and the broad lines disappeared making it a Type 1.9 $-$ 2 AGN \citep{2014ApJ...796..134D}. The MUSE spectra obtained by \cite{2019MNRAS.486..123R} recently in October 2017 showed strong broad Balmer component confirming the re-appearance of the BLR and transition of Mrk 590 from Type 2 to Type 1. Our Subaru spectrum obtained in October 2018 shows both H$\beta$ and H$\alpha$ broad-lines with FWHM of $10390\pm1718$ km s$^{-1}$ and $6478\pm240$ km s$^{-1}$, respectively. This reconfirms that Mrk 590 is currently in the Type 1 state. Using the measured H$\alpha$ BLR size of $21.44^{+1.49}_{-2.11}$ light-days and the FWHM of the H$\alpha$ line, the black hole mass is found to be $1.96^{+0.15}_{-0.21}\times 10^8 M_{\odot}$ using virial relation and a scale factor of 1.12. The black hole mass from the single-epoch spectrum is found to be $(1.59\pm 0.14)\times 10^8 M_{\odot}$ based on the H$\alpha$ line luminosity and the FWHM using the relation given by \citet{2015ApJ...801...38W}. Both black hole masses are consistent with each other. Comparing the black hole mass of $(4.75\pm0.74) \times 10^7$ M$\odot$ measured by \cite{2004ApJ...613..682P}, our measurement is a factor $\sim 4$ larger, which is due to the increase in the line width of the broad Balmer component by a factor of $\sim2$. \citet{2014A&A...568A..36P} found that symmetric cutting of emission line wings by the narrow-band filter may lead to over estimation of the BLR size by less than 5 percent. Taking this also into account, our estimated BLR size and black hole mass may have additional uncertainty of not more than 5 percent.





\section{SUMMARY}\label{sec:summary}

We carried out photometric reverberation mapping of Mrk 590 using broad $B$ and $R$ filters, as well as narrow H$\alpha$ and S II bands. The narrow S II-filter covers the redshifted H$\alpha$ line allowing us to construct H$\alpha$ line light curve and estimate the size of the BLR in Mrk 590 through cross-correlation analysis. Our main results are summarized below. 

\begin{enumerate}

\item
A significant contribution of the host-galaxy was found in all bands. This contribution was removed using two dimensional image decomposition of the stacked images. The host-corrected luminosity of the AGN in $B$, $R$, H$\alpha$ and S II bands are $\log L_{\mathrm{AGN}}$ = $43.36 \pm 0.01$, $43.53 \pm 0.01$, $42.81 \pm 0.06$ and $43.71 \pm 0.01$ erg s$^{-1}$, respectively.

\item
All the light curves show significant flux variations, which enabled us to carry out time series analysis to estimate the BLR size of Mrk 590.  We constructed $\mathrm{H\alpha}$ line light curve using two different methods. We corrected the continuum contamination in the narrow S II-band (1) using the observed narrow $\mathrm{H\alpha}$-band data which traces the continuum close to the S II-band and (2) considering a fixed PL slope of $\alpha=1/3$ to the observed B-band data. We found no difference in the measured lags between method 1 and 2 both using CCF and JAVELIN analysis. 

\item The broad $R$-band data also contains the H$\alpha$ line flux. We removed the continuum flux in the observed $R$-band by extrapolating the continuum measured in the $B$-band and assuming an $\alpha$ value of 1/3. We found a lag of $\sim$16 days between the $B$-band light curve and the continuum subtracted $R$-band light curve ($\mathrm{H\alpha}^{\prime\prime}$). The lower value of the lag is due to the larger contribution of the AGN continuum to the broad $R$-band compared to the narrow S II band.

\item Our estimated BLR size based on H$\alpha$ line in the rest frame of the source is  $21.44^{+1.49}_{-2.11}$ days, which is equivalent to 0.018 pc. The H$\alpha$ BLR size of Mrk 590 is consistent with the best-fit size-luminosity relation of AGN. 

\item The Subaru spectrum shows strong H$\beta$ and H$\alpha$ lines with FWHM of $6478\pm240$ and $10390\pm1718$ km s$^{-1}$, respectively, suggesting the transition of Mrk 590 from Type 2 (in 2006 $-$ 2013) to Type 1 (in 2017 $-$ 2019). The black hole mass measured from the single-epoch Subaru spectrum is found to be $1.96^{+0.15}_{-0.21}\times 10^8 M_{\odot}$.

\end{enumerate}

\section*{Acknowledgements}

 We thank the anonymous referee for her/his critical comments that led to the improvement of the manuscript. AKM and RS thank the National Academy of Sciences, India (NASI), Prayagraj for funding and Director, IIA for hosting and providing infrastructural support to this project. AKM also thank the Humboldt foundataion, Germany for the funding to visit ESO, Germany, where part of the reduction of the observed data and analysis were carried out. This research is based in part on data collected at Subaru Telescope,
which is operated by the National Astronomical Observatory of Japan.
We are honored and grateful for the opportunity of observing the Universe from Maunakea, which has the cultural, historical and natural significance in Hawaii. 


\section*{DATA AVAILABILITY}
The data underlying this article are available in the article
and in its online supplementary material.


\clearpage


\bsp	
\label{lastpage}
\end{document}